\documentclass[twocolumn,english,aps,prd,nofootinbib,hidelinks]{revtex4-2}
\usepackage{lmodern}
\usepackage[LGR,T1]{fontenc}
\usepackage[latin9]{inputenc}
\usepackage{xcolor}
\usepackage{babel}
\usepackage{amsmath}
\usepackage{amssymb}
\usepackage{graphicx}
\usepackage[unicode=true]
 {hyperref}

\makeatletter

\DeclareRobustCommand{\greektext}{%
  \fontencoding{LGR}\selectfont\def\encodingdefault{LGR}}
\DeclareRobustCommand{\textgreek}[1]{\leavevmode{\greektext #1}}
\ProvideTextCommand{\~}{LGR}[1]{\char126#1}

\providecommand{\tabularnewline}{\\}

\usepackage{enumerate}

\makeatother

\begin{document}
\title{New analysis of SNeIa Pantheon Catalog: $\,$Variable speed of light
as \vskip1pt an alternative to dark energy}
\author{Hoang Ky Nguyen$\,$}
\email[\ ]{hoang.nguyen@ubbcluj.ro}

\affiliation{{\vskip2pt}Department of Physics, Babe\c{s}-Bolyai University, Cluj-Napoca
400084, Romania{\vskip1pt}Institute for Interdisciplinary Research
in Science and Education,~\linebreak{}
 ICISE, Quy Nhon 55121, Vietnam}

\date{April 7, 2025}
\begin{abstract}
\vskip2ptIn \textcolor{blue}{\href{https://arxiv.org/abs/astro-ph/0304237}{A\&A {\bf 412}, 35 (2003)}}
Blanchard, Douspis, Rowan-Robinson, and Sarkar (BDRS) slightly modified
the primordial fluctuation spectrum and produced an excellent fit
to WMAP's CMB power spectrum for an Einstein--de Sitter (EdS) universe,
\emph{bypassing} dark energy. Curiously, they obtained a Hubble value
of $H_{0}\approx46$, in sharp conflict with the canonical range $\sim67\text{--}73$.
However, we will demonstrate that the \emph{reduced} value of $H_{0}\approx46$
achieved by BDRS is fully compatible with the use of \emph{variable
speed of light} in analyzing the late-time cosmic acceleration observed
in Type Ia supernovae (SNeIa). In \textcolor{blue}{\href{https://www.sciencedirect.com/science/article/pii/S0370269325001170}{Phys.\ Lett.\ B {\bf 862} (2025) 139357}}
we considered a generic class of scale-invariant actions that allow
matter to couple non-minimally with gravity via a dilaton field $\chi$.
We discovered a hidden aspect of these actions: $\:$the dynamics
of the dilaton can induce a variation in the speed of light $c$ as
$c\propto\chi^{1/2}$, thereby causing $c$ to vary alongside $\chi$
across spacetime. For an EdS universe with varying $c$, besides the
effects of cosmic expansion, light waves emitted from distant SNeIa
are further subject to\emph{ a refraction effect, which alters the
Lema\^itre redshift relation to $1+z=a^{-3/2}$.} Based on this new
formula, we achieve a fit to the SNeIa Pantheon Catalog exceeding
the quality of the $\Lambda$CDM model. Crucially, our approach does
\emph{not} require dark energy and produces $H_{0}=47.2\pm0.4$ (95\%
CL) in strong alignment with the BDRS finding of $H_{0}\approx46$.
The reduction in $H_{0}$ in our work, compared with the canonical
range $\sim67\text{--}73$, \emph{arises due to the $3/2$--exponent}
in the modified Lema\^itre redshift formula. Hence, BDRS's analysis
of the (early-time) CMB power spectrum and our variable-$c$ analysis
of the (late-time) Hubble diagram of SNeIa fully agree on two counts:\linebreak
(i) \emph{the dark energy hypothesis is avoided}, and (ii) \emph{$H_{0}$
is reduced to $\sim47$}, which also yields an age $t_{0}=2/(3H_{0})=13.8\,$Gy
for an EdS universe, without requiring dark energy. Most importantly,
we will demonstrate that \emph{the late-time acceleration can be attributed
to the declining speed of light in an expanding EdS universe}, rather
than to a dark energy component.
\end{abstract}
\maketitle

\section{\label{sec:Motivation}Motivation}

The $\Lambda$CDM model serves as the standard framework for modern
cosmology, efficiently accounting for a wide range of astronomical
observations. While the model is widely regarded as successful, it
faces significant challenges \citep{Perivolaropoulos-2022}. Notably,
ongoing tensions in the determination of the Hubble constant $H_{0}$
and the amplitude of matter fluctuations $\sigma_{8}$ raise questions
about the underpinning principles of the model \citep{diValentino-2021}.
Moreover, an integral component of this model is dark energy (DE),
which constitutes approximately 70\% of the total energy budget of
the universe. The nature of DE itself---along with its fine-tuning
and coincidence problems---poses profound challenges both in cosmology
and in the broader context of field theories \citep{Bull}.\vskip4pt

In 2003, Blanchard, Douspis, Rowan-Robinson, and Sarkar (BDRS) proposed
a novel approach to mitigate the need for DE when analyzing the cosmic
microwave background (CMB) power spectrum \citep{BDRS-2003}. They
relied on the Einstein--de Sitter universe, which corresponds to
the flat $\Lambda$CDM model with $\Omega_{\Lambda}=0$. Instead of
the conventional single-power primordial fluctuation spectrum, $P(k)\!=\!Ak^{n}$,
they employed a double-power form\vskip-10pt
\begin{equation}
P(k)=\begin{cases}
A_{1}\,k^{n_{1}} & k\leqslant k_{*}\\
A_{2}\,k^{n_{2}} & k\geqslant k_{*}
\end{cases}\label{eq:double-power}
\end{equation}
with continuity imposed across the breakpoint $k_{*}$. Remarkably,
this modest modification produced an excellent fit to the CMB power
spectrum without invoking DE. In \citep{Hunt-2007}, Hunt and Sarkar
further developed a supergravity-based inflation scenario to validate
the double-power form given in Eq.$\ $\eqref{eq:double-power} and
also attained an excellent fit while avoiding DE. The works by BDRS
and the Hunt--Sarkar team---\emph{if correct}---would seriously
undermine the viability of the DE hypothesis.\vskip6pt

Surprisingly, the fit by BDRS yielded a new value of $H_{0}\approx46$,
while the fit by Hunt and Sarkar produced a comparable value of $H_{0}\approx44$.
Obviously, these values are at odds with the value of $H_{0}\sim70$
derived from the Hubble diagram of Type Ia supernovae (SNeIa), based
on the $\Lambda$CDM model with $\Omega_{\Lambda}\approx0.7$. Since
DE has been regarded as the driving force of the late-time cosmic
acceleration, interest in the works by BDRS and the Hunt--Sarkar
team has largely diminished in favor of the standard $\Lambda$CDM
model.\vskip7pt

Amid this backdrop, we will reexamine the SNeIa data in the context
of a cosmology that supports a varying speed of light $c$ on an expanding
EdS cosmic background, rather than the $\Lambda$CDM model. Recently,
the theoretical foundation for a varying $c$ in spacetime has been
derived, using a class of scale-invariant actions that enable non-minimal
coupling of matter with gravity via a dilaton field \citep{VSL2024-dilaton,VSL2024-1}.
In cosmology, a varying $c$ should impact the propagation of light
rays from distant SNeIa to an Earth-based observer, fundamentally
altering the distance-vs-redshift relationship. This modification
necessitates a \emph{re-evaluation of the $H_{0}$ value derived from
the Hubble diagram of SNeIa}, potentially replacing the canonical
value $\sim70$ that relies on the $\Lambda$CDM model.\vskip4pt

The purpose of our paper is two-fold: (i) to investigate the viability
of the variable speed of light (VSL) theory developed in Refs. \citep{VSL2024-dilaton,VSL2024-1}
in accounting for the late-time cosmic acceleration while bypassing
DE, and (ii) to determine whether---and how---the finding of $H_{0}\approx46$
by BDRS for the CMB can be reconciled with our reexamination of SNeIa
in the VSL context.\vskip8pt

Our paper is organized into four major parts:\vskip4pt

\textbf{\emph{$*$ Foundation of VSL:}} Section \ref{sec:Generating-VSL}
covers the history of VSL and provides a recap of our mechanism for
generating varying $c$, as presented in \citep{VSL2024-dilaton,VSL2024-1}.\vskip4pt

\textbf{\emph{$*$ Cosmography of VSL:}} Sections \ref{sec:Impacts-of-VSL-in-EdS}
and \ref{sec:Refractive} prepare the foundational material necessary
for cosmography in the presence of varying $c$. Section \ref{sec:Modifications-all-formulae}
develops various \emph{modified} redshift relations---Lema\^itre,
distance vs. $z$, and luminosity distance vs. $z$---by incorporating
variations in the speed of light. Importantly, we derive \emph{a modified
Hubble law, applicable to our VSL scheme}.\vskip4pt

\textbf{\emph{$*$ Cosmology of VSL:}} Section \ref{sec:Reanalysis-Pantheon}
presents our analysis of the Combined Pantheon Sample of SNeIa data
using our modified luminosity distance vs. redshift formula derived
in the preceding sections.\vskip4pt

\textbf{\emph{$*$ Consequences of VSL:}} We aim for four objectives:
(i) Section \ref{sec:VSL-as-alternative} presents a new interpretation
of the accelerating expansion through varying $c$ instead of DE;
(ii) Section \ref{sec:BDRS} revisits BDRS's analysis of the CMB power
spectrum without requiring DE and reconcile it with our findings from
the VSL-based analysis of SNeIa; (iii) Section \ref{sec:Resolve-age-prob}
resolves the age problem without using DE; and (iv) Section \ref{sec:Resolving-H0-tension}
offers a potential resolution to the $H_{0}$ tension from an astronomical
origin while avoiding dynamical DE.\vskip4pt

Section \ref{sec:Summary} discusses and summarizes our findings,
and the appendices contain technical supplements.

\vspace{-.15cm}

\section{\label{sec:Generating-VSL}A new mechanism to generate varying $\large{\textbf{c}}$
from dilaton dynamics}

The variability in the speed of light was first recognized by Einstein
in 1911 during his pursuit for a generally covariant theory of gravitation,
which ultimately culminated in the theory of General Relativity (GR)
in 1915. In Ref.$\ $\citep{Einstein-1911} he explicitly allowed
the gravitational field $\Phi$ to influence the value of $c$ in
spacetime. In particular, he proposed that $c=c_{0}\left(1+\Phi/c^{2}\right)$,
where $c_{0}$ is the speed of light at a reference point where $\Phi$
vanishes. Notably, he conceived this radical idea six years \emph{after}
his formulation of Special Relativity (SR). As Einstein emphasized
in \citep{Einstein-1912-a,Einstein-1912-b}, a variation in $c$ does
not contradict the principle of the constancy of $c$ under Lorentz
transformations, an underpinning requirement of SR. This is because
Lorentz invariance, confirmed by the Michelson-Morley (MM) experiment,
is only required to hold in \emph{local} inertial frames and does
not necessitate its global validity in \emph{curved} spacetimes. More
concretely, in a region vicinity to a given point $x^{*}$, the tangent
frames to the spacetime manifold possess the Lorentz symmetry with
a common value of $c$ \emph{applicable only to that region}. Yet,
in a spacetime influenced by a gravitational field, different regions
can---in principle---correspond to different values of $c$. Utilizing
the language of Riemannian geometry, the speed of light can be promoted
to a scalar field: while $c$ is an \emph{invariant} (i.e., unaffected
upon diffeomorphism), it can nonetheless be \emph{position-dependent},
viz. $c(x^{*})$.\vskip4pt

Einstein's pioneering concept of VSL, nevertheless, was quickly overshadowed
by the success of his GR and subsequently fell into dormancy for several
decades. The variability of $c$ was briefly rediscovered by Dicke
in 1957 \citep{Dicke-1957}, prior to his own development of Brans--Dicke
gravity \citep{BransDicke-1961}, which instead allowed Newton's gravitational
constant $G$ to vary. In the 1990s, the idea of VSL was independently
revived by Moffat \citep{Moffat-1993} and by Albrecht and Magueijo
\citep{Albrecht-1998} in the context of early-time cosmology. Their
proposals aimed to resolve the horizon puzzle while avoiding the need
for cosmic inflation. Since then, several researchers actively explore
various aspects of VSL \citep{Zhang-2014,Qi-2014,Ravanpak-2017,Barrow-2000,Bassett-2000,Drummond-1980,Drummond-1999,Liberati-2000,Magueijo-2002,Magueijo-2003,Magueijo-2008,Novello-1989,Volovik-2023,Ellis-2005,Ellis-2007,Zou-2018,Zhang-2024,Wang-2019,Uzan-2011,Uzan-2003,Santos-2024,Salzano-2017b,Salzano-2017a,Salzano-2016b,Salzano-2016a,Salzano-2015,Rodrigues-2022,Pan-2020,Mukherjee-2024,Moffat-2016,Mendonca-2021,Martins-2017,Magueijo-2000,Liu-2023,Liu-2021,Lee-2023b,Lee-2023a,Lee-2021b,Lee-2021a,Lang-2018,HWAC-2018,HAWC-2020,Gupta-2021,Gupta-2020,Franzmann-2017,Eaves-2022,Cuzinatto-2023,Cuzinatto-2022,Cruz-2018,Cruz-2012,Costa-2019,Colaco-2022,Clayton-2002,Clayton-2000,Clayton-1999,Cao-2018,Cao-2017,Cai-2016,Barrow-1999b,Barrow-1999a,Barrow-1998b,Barrow-1998a,Balcerzak-2014b,Balcerzak-2014a,Avelino-1999,Agrawal-2021,Abdo-2009,Salzano-2016c,Liu-2018,Xu-2016a,Xu-2016b,Zhu-2021,Avelino-2000,Buchalter-2004}.\vskip4pt

In a recent report \citep{VSL2024-dilaton}, we considered a scale-invariant
action that facilitates non-minimal coupling of matter with gravity
via a dilaton field $\chi$. We uncovered a hidden mechanism that
induces a dependence of $c$ and $\hbar$ on the dilaton field $\chi$,
thereby causing $c$ and $\hbar$ to vary alongside $\chi$ in spacetime.
Below is a recap of our mechanism.

\subsection*{The essence of our VSL mechanism}

Let us consider a prototype action:
\begin{align}
\mathcal{S} & =\int d^{4}x\sqrt{-g}\,\Bigl[\mathcal{L}_{\text{grav}}+\mathcal{L}_{\text{mat}}\Bigr]\label{eq:S2}\\
\mathcal{L}_{\text{grav}} & =\chi^{2}\,\mathcal{R}-4\omega g^{\mu\nu}\partial_{\mu}\chi\partial_{\nu}\chi\label{eq:L-BD}\\
\mathcal{L}_{\text{mat}} & =i\,\bar{\psi}\gamma^{\mu}\nabla_{\mu}\psi+\sqrt{\alpha}\,\bar{\psi}\gamma^{\mu}A_{\mu}\psi+\mu\,\chi\,\bar{\psi}\psi-\frac{1}{4}F_{\mu\nu}F^{\mu\nu}\label{eq:L-chi-QED}
\end{align}
The gravitational sector $\mathcal{L}_{\text{grav}}$ is equivalent
to the well-known Brans--Dicke theory, $\mathcal{L}_{\text{BD}}=\phi\,\mathcal{R}-\frac{\omega}{\phi}g^{\mu\nu}\partial_{\mu}\phi\partial_{\nu}\phi$,
upon substituting $\phi:=\chi^{2}$ \citep{BransDicke-1961}. The
matter Lagrangian $\mathcal{L}_{\text{mat}}$ describes quantum electrodynamics
(QED) for an electron field $\psi$, coupled with an electromagnetic
field $A_{\mu}$ (with the field tensor defined as $F_{\mu\nu}:=\partial_{\mu}A_{\nu}-\partial_{\nu}A_{\mu}$)
and embedded in a curved spacetime characterized by the metric $g^{\mu\nu}$.
The Dirac gamma matrices satisfy $\gamma^{\mu}\gamma^{\nu}+\gamma^{\nu}\gamma^{\mu}=2g^{\mu\nu}$,
and the spacetime covariant derivative $\nabla_{\mu}$ acts on the
spinor via vierbein and spin connection. However, the electron field
couples \emph{non-minimally} with gravity via the dilaton field, viz.
$\chi\,\bar{\psi}\psi$. \vskip4pt

All parameters $\alpha$, $\mu$, and $\omega$ are dimensionless.
The full action is scale invariant, viz. unchanged under the global
Weyl rescaling
\begin{equation}
g_{\mu\nu}\rightarrow\Omega^{2}g_{\mu\nu};\ \chi\rightarrow\Omega^{-1}\chi;\ \psi\rightarrow\Omega^{-\frac{3}{2}}\psi;\ A_{\mu}\rightarrow A_{\mu}
\end{equation}
It has been established in \citep{Ferreira-2016,Blas-2011} that a
scale-invariant action, such as the one described in Eqs. \eqref{eq:S1}--\eqref{eq:L-QED},
can evade observational constraints on the fifth force.\vskip8pt

Next, let us revisit the `canonical' QED action for an electron field
$\psi$ carrying a $U(1)$ gauge charge $e$ and inertial mass $m$,
coupled with an electromagnetic field $A_{\mu}$ and embedded in an
Einstein--Hilbert spacetime

\begin{align}
\mathcal{S}_{0} & =\int d^{4}x\sqrt{-g}\,\Bigl[\mathcal{L}_{\text{EH}}+\mathcal{L}_{\text{QED}}\Bigr]\label{eq:S1}\\
\mathcal{L}_{\text{EH}} & =\frac{c^{3}}{16\pi\hbar G}\,\mathcal{R}\label{eq:L-EH}\\
\mathcal{L}_{\text{QED}} & =i\,\bar{\psi}\gamma^{\mu}\nabla_{\mu}\psi+\frac{e}{\sqrt{\hbar c}}\bar{\psi}\gamma^{\mu}A_{\mu}\psi+m\frac{c}{\hbar}\,\bar{\psi}\psi-\frac{1}{4}F_{\mu\nu}F^{\mu\nu}\label{eq:L-QED}
\end{align}
In these expressions, the quantum of action $\hbar$, speed of light
$c$, and Newton's gravitational parameter $G$ are explicitly restored.\vskip4pt

Excluding the kinetic term $g^{\mu\nu}\partial_{\mu}\chi\partial_{\nu}\chi$
of the dilaton in Eq. \eqref{eq:L-BD}, the action $\mathcal{S}$
can be brought into the form $\mathcal{S}_{0}$ via the following
identification:
\begin{equation}
\frac{c^{3}}{16\pi\hbar G}:=\chi^{2};\ \ \ \frac{e}{\sqrt{\hbar c}}:=\sqrt{\alpha};\ \ \ m\frac{c}{\hbar}:=\mu\,\chi\label{eq:identification}
\end{equation}
These identities link the charge $e$ and inertial mass $m$ of the
electron with the three `fundamental constants' $c$, $\hbar$, and
$G$, as well as the dilaton field $\chi$.\vskip8pt

To proceed, we \emph{require} that the intrinsic properties of the
electron---namely, its charge $e$ and \emph{inertial} mass $m$---remain
independent of the background spacetime, particularly the dilation
field $\chi$ which belongs to the gravitational sector. Consequently,
based on the last two identities of Eq. \eqref{eq:identification},
both the speed of light $c$ and the quantum of action $\hbar$ must
be treated as scalar fields related to $\chi$. The following assignments
capture this relationship:
\begin{equation}
c_{\chi}:=\hat{c}\ \Bigl(\frac{\chi}{\hat{\chi}}\Bigr)^{1/2};\ \ \ \hbar_{\chi}:=\hat{\hbar}\ \Bigl(\frac{\chi}{\hat{\chi}}\Bigr)^{-1/2}\label{eq:scheme2-assign}
\end{equation}
Here, the subscript $\chi$ signifies the dependence of $c$ and $\hbar$
on $\chi$, while $\hat{c}$ and $\hat{\hbar}$ represent the values
of $c_{\chi}$ and $\hbar_{\chi}$ at a reference point where $\chi=\hat{\chi}$
(with $\hat{\chi}\neq0$). It is straightforward to derive from Eqs.
\eqref{eq:identification} and \eqref{eq:scheme2-assign} that
\begin{equation}
e=\bigl(\alpha\hat{\hbar}\hat{c}\bigr)^{1/2};\ \ \ m=\frac{\mu\hat{\hbar}\hat{\chi}}{\hat{c}};\ \ \ G=\frac{\hat{c}^{3}}{16\pi\hat{\hbar}\hat{\chi}^{2}}\label{eq:e-m-scheme2}
\end{equation}
This confirms that $e$ and $m$ are indeed constants \footnote{Note: When radiative corrections in the matter sector are taken into
account, $e$ and $m$ can `run' in the renormalization group flow
as functions of the momentum level at which they are measured.}. Furthermore, $G$ is also constant.\vskip4pt

As the dilaton $\chi$ varies in spacetime as a component of the gravitational
sector, the scalar fields $c_{\chi}$ and $\hbar_{\chi}$, defined
in Eq. \eqref{eq:scheme2-assign}, also vary in spacetime. Therefore,
\emph{the dynamics of the dilaton $\chi$ induces variations in $c_{\chi}$
and $\hbar_{\chi}$ on the spacetime manifold.}

\subsection*{Comments on Brans--Dicke's variable $G$}

Traditionally, Brans--Dicke (BD) gravity is associated with variable
Newton's gravitational constant $G$ \citep{BransDicke-1961}. It
should be noted that Brans and Dicke only allowed matter to couple
minimally with gravity, namely, through the 4--volume element $\sqrt{-g}$;
in this case, the matter action is not scale invariant. To achieve
scale invariance, matter must couple with gravity in a non-minimal
way, such as the Lagrangian given in Eq. \eqref{eq:L-chi-QED}. In
this case, \emph{if one presumes that $c$ and $\hbar$ are constants},
then the mass parameters of (massive) fields also become variable
\citep{Fujii-1982,Wetterich-1988a,Wetterich-1988b,Wetterich-2013a}.\vskip4pt

Indeed, under the assumption of constant $c$ and $\hbar$, Eq. \eqref{eq:identification}
readily produces
\begin{equation}
e=\bigl(\alpha\hbar c\bigr)^{1/2};\ \ \ m_{\chi}=\frac{\mu\hbar}{c}\,\chi;\ \ \ G_{\chi}=\frac{c^{3}}{16\pi\hbar}\,\chi^{-2}\label{eq:FW-scheme}
\end{equation}
Here, the subscript $\chi$ signifies the dependence of $m$ and $G$
on $\chi$. In \citep{VSL2024-dilaton}, we referred to these results
as ``the Fujii--Wetterich scheme'', since these authors appear to
be the first to report results (in \citep{Fujii-1982,Wetterich-1988a,Wetterich-1988b,Wetterich-2013a})
essentially equivalent to Eq. \eqref{eq:FW-scheme}. In this scheme,
while $m$ is associated with $\chi$, the charge $e$ remains independent
of $\chi$, rendering an \emph{unequal} treatment of $e$ and $m$.
Moreover, whereas $\chi$ affects the electron's mass per Eq. \eqref{eq:FW-scheme},
massless particles, such as photons, remain \emph{unaffected}.\vskip4pt

Our mechanism thus represents a significant departure from the variable
$G$ (and mass) approach. Importantly, it allows the dilaton $\chi$---through
its influence on the speed of light $c_{\chi}$ and quantum of action
$\hbar_{\chi}$---to govern the propagation and quantization of \emph{all
fields}---viz. electron and photon---on a universal and equal basis.\vskip4pt

While both approaches---(i) variable $G$ and $m$ versus (ii) variable
$c$ and $\hbar$---are mathematically permissible, they are \emph{not
physically equivalent} \citep{VSL2024-dilaton}, and the validity
of each approach should be determined through empirical evidence,
including predictions, experiments and observations.\vskip4pt

Our mechanism leads to a direct and immediate consequence in cosmology,
however. Specifically, the aspect of our mechanism where the dynamical
$\chi$ induces a variation in $c_{\chi}$, which in turn governs
\emph{massless} field (viz. the light quanta) has significant implications.
A varying $c$ influences the propagation of light rays emitted from
distant sources toward an Earth-based observer, thereby affecting
the Hubble diagram of these light sources, particularly for SNeIa.
This intuition serves as the underpinning for the analysis presented
in the remainder of this paper.

\subsection*{Scaling properties of length, time, and energy}

In \citep{VSL2024-dilaton} we further deduced that at a given point
$x^{*}$, the prevailing value of the dilaton $\chi(x^{*})$ determines
the lengthscale, timescale, and energy scale for physical processes
occurring at that point. The lengthscale $l$ and energy scale $E$
are dependent on $\chi$ as follows
\begin{equation}
l\propto\chi^{-1};\ \ \ E\propto\chi\,.\label{eq:length-vs-chi}
\end{equation}
However, the most important outcome is that the timescale $\tau$
behaves in an \emph{anisotropic} fashion, as
\begin{equation}
\tau\propto\chi^{-3/2}
\end{equation}
or, equivalently
\begin{equation}
\tau\propto l^{\,3/2}\,.\label{eq:tau-vs-length}
\end{equation}
This leads to a novel time dilation effect induced by the dilaton,
representing a concrete prediction of our mechanism. Moreover, the
$3/2$--exponent in this time scaling law plays a crucial role in
the Hubble diagram of SNeIa, as we will explore in the following sections.
\vskip8pt

A detailed exposition of our mechanism and the new time dilation effect
is presented in Ref. \citep{VSL2024-dilaton}.

\section{\label{sec:Impacts-of-VSL-in-EdS}Impacts of varying $\large{\textbf{c}}$
in an Einstein--de Sitter universe}

In a cosmology accommodating VSL, as a lightwave travels from a distant
SNeIa toward an Earth-based observer, \emph{a varying speed of light
along its trajectory induces a refraction effect} akin to that experienced
by a physical wave traveling through an inhomogenous medium with varying
wave speed. The alteration of the wavelength results in a new set
of cosmographic formulae, including a \emph{modified} Hubble law and
a\emph{ modified }relationship between redshift and luminosity distance.

\subsection{\label{subsec:Drawback}A drawback in previous VSL analyses of SNeIa}

It is important to note that since the revival of VSL by Moffat and
the Albrecht--Magueijo team in the 1990s, several authors have applied
VSL to late-time cosmology, particularly in the analysis of the Hubble
diagram of SNeIa. However, these attempts have not met with much success
\citep{Zhang-2014,Qi-2014,Ravanpak-2017,Salzano-2016a,Salzano-2016b,Barrow-1998a,Barrow-1998b}.
A common theme among these analyses is the assumption that $c$ varies
as a function of the global cosmic factor $a$ of the Friedmann--Lema\^itre--Roberson--Walker
(FLRW) metric (e.g., in the form $c\propto a^{-\zeta}$ first proposed
by Barrow \citep{Barrow-1998a}). These works generally conclude that,
despite the dependence of $c$ on $a$, VSL does not alter the classic
Lema\^itre redshift formula $1+z=a^{-1}$ and, therefore, cannot
play any role in the Hubble diagram of SNeIa. However, upon closer
scrutiny into these works, we identify a significant oversight: they
implicitly assumed that $c$ is a function solely of cosmic time $t$,\linebreak
through the dependence of $a$ on $t$ in the FLRW metric. This assumption
is not valid in our VSL framework, where $c$---through its dependence
on the dilaton field $\chi$---varies in both \emph{space} and time,
rather than time alone. In this section, as well as Sections \ref{sec:Refractive}
and \ref{sec:Modifications-all-formulae}, we will demonstrate that
\emph{the variation of $c$ as a function of the dilaton field $\chi$,
rather than merely as a function of the cosmic factor (viz. $a$)
as assumed in prior VSL works, fundamentally alters the Lema\^itre
redshift formula and necessitates a re-analysis of SNeIa data.}

\subsection{\label{subsec:Modified-RW}The modified FLRW metric}

The FLRW metric for the isotropic and homogeneous intergalactic space
reads
\begin{align}
ds^{2} & =c^{2}dt^{2}-a^{2}(t)\,\left[\frac{dr^{2}}{1-\kappa r^{2}}+r^{2}d\Omega^{2}\right]\label{eq:RW}
\end{align}
where $a(t)$ is the global cosmic scale factor that evolves with
cosmic time $t$.\vskip4pt

Our goal is to investigate whether an Einstein--de Sitter universe,
when supplemented with a varying $c$, can account for the Hubble
diagram of SNeIa as provided by the Pantheon Catalog. We will make
three working assumptions:\vskip8pt

\textbf{\emph{Assumption \#1:}} The FLRW universe is spatially flat,
corresponding to $\kappa=0$. There is robust observational evidence
supporting this assumption.\vskip8pt

\textbf{\emph{Assumption \#2:}} The cosmic scale factor evolves as
\begin{equation}
a=a_{0}\left(\text{\ensuremath{\frac{t}{t_{0}}}}\right)^{2/3}\label{eq:a-vs-t}
\end{equation}
\emph{Justification: }In our VSL mechanism, the timescale $\tau$
and lengthscale $l$ of a given physical process are related by $\tau\propto l^{\,3/2}$,
as expressed in Eq. \eqref{eq:tau-vs-length}. Regarding the evolution
of the FLRW metric, its timescale and lengthscale can be identified
with $t$ and $a$, respectively. The growth law given in Eq. \eqref{eq:a-vs-t}
is therefore justified. Note: This growth is identical to the evolution
of an EdS universe, viz. a spatially flat, expanding universe consisting
solely of matter, with no contribution from DE or a cosmological constant.\vskip8pt

\textbf{\emph{Assumption \#3:}} The dilaton field in the cosmic background
depends on the cosmic factor in the form
\begin{equation}
\chi\propto a^{-1}\,.\label{eq:chi-vs-a}
\end{equation}
\emph{Justification: }In our VSL mechanism, the lengthscale of a given
physical process is inversely proportional to the dilaton field, per
Eq. \eqref{eq:length-vs-chi}. Given that the cosmic factor $a$ plays
the role of the lengthscale for the FLRW metric, the dependency expressed
in Eq. \eqref{eq:chi-vs-a} is therefore justified. \vskip8pt

Combining Eqs. \eqref{eq:scheme2-assign} and \eqref{eq:chi-vs-a}
then renders $c\propto a^{-1/2}$, or more explicitly
\begin{equation}
c=c_{0}\left(\frac{a}{a_{0}}\right)^{-1/2}\label{eq:c-vs-a}
\end{equation}
Here, $a_{0}$ is the current cosmic scale factor (often set equal
1), and $c_{0}$ is the speed of light measured at our current time
in the intergalactic space. We should emphasize that the value of
$c_{0}$ is not identical with the one measured \emph{inside the Milky
Way}, which is equal to $300,000$ km/s. This is because the Milky
Way is a gravitationally bound structure whereas the intergalactic
space is regions subject to cosmic expansion. This issue will be explained
in Section \ref{sec:Refractive}.\vskip8pt

Combining Eqs. \eqref{eq:RW} and \eqref{eq:c-vs-a}, and setting
$\kappa=0$, we then obtain the \emph{modified} FLRW metric
\begin{align}
ds^{2} & =\frac{c_{0}^{2}a_{0}}{a(t)}\,dt^{2}-a^{2}(t)\,\left[dr^{2}+r^{2}d\Omega^{2}\right]\label{eq:modified-RW}
\end{align}
which describes\emph{ an EdS universe with a declining speed of light,
per $c\propto a^{-1/2}$.}

\subsection{\label{subsec:Frequency-shift}Frequency shift}

For the modified FLRW metric derived above, the null geodesic ($ds^{2}=0$)
for a lightwave traveling from a distant emitter toward Earth (viz.
$d\Omega=0$) is
\begin{equation}
c_{0}\,a_{0}^{1/2}\frac{dt}{a^{3/2}(t)}=dr\label{eq:null_geodesic}
\end{equation}
Hereafter, we will use the subscripts ``\emph{em}'' and ``\emph{ob}''
for ``emission'' and ``observation'' respectively. Denote $t_{em}$
and $t_{ob}$ the emission and observation time points of the lightwave,
and $r_{em}$ the co-moving distance of the emitter from Earth. From
\eqref{eq:null_geodesic}, we have:
\begin{equation}
c_{0}\,a_{0}^{1/2}\int_{t_{em}}^{t_{ob}}\frac{dt}{a^{3/2}(t)}=r_{em}
\end{equation}
The next wavecrest to leave the emitter at $t_{em}+\delta t_{em}$
and arrive at Earth at $t_{ob}+\delta t_{ob}$ satisfies:
\begin{equation}
c_{0}\,a_{0}^{1/2}\int_{t_{em}+\delta t_{em}}^{t_{\text{ob}}+\delta t_{ob}}\frac{dt}{a^{3/2}(t)}=r_{em}
\end{equation}
Subtracting these two equations yields:
\begin{equation}
\frac{\delta t_{ob}}{a^{3/2}(t_{ob})}=\frac{\delta t_{em}}{a^{3/2}(t_{em})}
\end{equation}
which leads to the ratio between the emitted frequency and the observed
frequency:
\begin{equation}
\frac{\nu_{ob}}{\nu_{em}}=\frac{\delta t_{em}}{\delta t_{ob}}=\frac{a^{3/2}(t_{em})}{a^{3/2}(t_{ob})}=\left(\frac{a_{em}}{a_{ob}}\right)^{3/2}\label{eq:freq_shift}
\end{equation}
This contrasts with the standard relation, $\frac{\nu_{ob}}{\nu_{em}}=\frac{a_{em}}{a_{ob}}$.
To derive a Lema\^itre formula applicable for VSL, further consideration
is needed. This task will be carried out in the next section.

\section{\label{sec:Refractive}Impacts of varying $\large{\textbf{c}}$ across
boundaries of galaxies}

This section presents the pivotal elements that enable the $3/2$-exponent
in the frequency ratio, as expressed in Eq. \eqref{eq:freq_shift},
to manifest in observations.
\begin{figure*}[!t]
\noindent \begin{centering}
\hskip-10pt\includegraphics[scale=0.74]{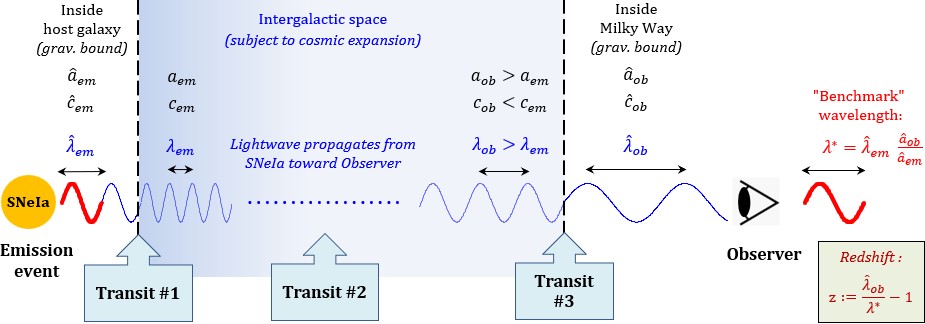}\vspace{0.25cm}
\par\end{centering}
\caption{A lightwave from an SNeIa explosion (shown on the far left) makes
three transits to reach the Earth-based astronomer (shown on the far
right). During Transit \#1, the lightwave exits the (gravitationally-bound)
host galaxy to enter the surrounding intergalactic region; the wavecrest
gets compressed as light speed decreases at this juncture. During
Transit \#2, the lightwave travels in the intergalactic space which
undergoes cosmic expansion; accordingly, the wavecrest expands. During
Transit \#3, the lightwave enters the (gravitationally-bound) Milky
Way; the wavecrest expands further as light speed increases at this
juncture. The Earth-based astronomer measures the wavelength $\hat{\lambda}_{ob}$
and compares it with the ``benchmark'' wavelength $\lambda^{*}$ (see
text for explanation) to calculate the redshift $z$ (shown in the
lower right corner box).}

\label{fig:wavetrain-full}
\end{figure*}

\subsection{The loss of validity of Lema\^itre formula}

Let us first revisit the drawback in previous VSL works alluded to
in Section \ref{subsec:Drawback}. The frequency ratio given by Eq.
\eqref{eq:freq_shift} can be converted into the wavelength ratio
\begin{equation}
\frac{\lambda_{ob}}{\lambda_{em}}=\frac{c_{ob}}{c_{em}}.\frac{\nu_{em}}{\nu_{ob}}=\frac{a_{ob}^{-1/2}}{a_{em}^{-1/2}}.\frac{a_{ob}^{3/2}}{a_{em}^{3/2}}=\frac{a_{ob}}{a_{em}}\label{eq:transit2}
\end{equation}
This expression is exactly \emph{identical} to that in standard cosmology,
viz. where $c$ is non-varying. At first, it may seem tempting to
relate the redshift $z$ with $\frac{\lambda_{ob}-\lambda_{em}}{\lambda_{em}}$,
namely
\begin{equation}
1+z\stackrel{?}{=}\frac{\lambda_{ob}}{\lambda_{em}}=\frac{a_{ob}}{a_{em}}=a^{-1}\label{eq:z-vs-a-false}
\end{equation}
in which $a_{ob}$ is set equal 1 and $a_{em}$ is denoted as $a$.
In Refs. \citep{Zhang-2014,Qi-2014,Ravanpak-2017,Barrow-1998b,Salzano-2016a,Salzano-2016b},
based on Eq. \eqref{eq:z-vs-a-false}, it was concluded that the classic
Lema\^itre redshift formula, $1+z=a^{-1}$, remained valid. Subsequently,
virtually all empirical VSL works continued using the classic Lema\^itre
formula to analyze the Hubble diagram of SNeIa.\vskip4pt

However, the formula in Eq. \eqref{eq:z-vs-a-false} is \emph{incorrect}.
One key reason is that $\lambda_{ob}$, representing the wavelength
in the intergalactic space enclosing the Milky Way, is \emph{not}
what the Earth-based astronomer directly measures. For the light wave
to reach the astronomer's telescope, it must pass through the gravitationally-bound
Milky Way, which has its own local scale $\hat{a}_{ob}$ \emph{differing}
from the current global cosmic scale because the matter-populated
Milky Way resists cosmic expansion. This crucial point will be clarified
shortly in the section below. In brief, a change in scale (from global
to local) across the boundary of the Milky Way induces a corresponding
change in the speed of light. This effect alters the wavelength from
$\lambda_{ob}$ to $\hat{\lambda}_{ob}$ which is then measured by
the astronomer.

\subsection{Refractive effect due to varying $\normalsize{\textbf{c}}$ across
boundaries of galaxies}

The global cosmic scale factor $a$ grows with comic time $t$, leading
to the stretching of wavelength of light from $\lambda_{em}$ to $\lambda_{ob}$.
However, the Solar System is not subject to cosmic expansion, which
is a crucial condition so that the Earth-based observer can detect
the redshift of distant emission sources. It is well understood that
if the Solar System expanded along with the intergalactic space, the
observer's instruments would also expand in sync with the wavelength
of the light ray emitted from a distant supernova, making the detection
of any redshift impossible. More generally, mature galaxies---those
hosting distant SNeIa, and the Milky Way where the Earth-based observer
resides, are \emph{gravitationally bound} and resist cosmic expansion.
Despite the expansion of intergalactic space, matured galaxies maintain
their relatively stable size primarily through gravitational attraction,
counterbalanced by the rotational motion of the matter within them.
Consequently, each galaxy has a stable local scale $\hat{a}$ that
remains relatively constant over time, despite increases in the global
scale $a$.\vskip4pt

As discussed in Section \ref{subsec:Modified-RW}, the dilaton field
$\chi$ in intergalactic space is inversely proportional to the global
scale factor $a$. Similarly, within a galaxy, the dilaton field is
inversely proportional to the galaxy's local scale $\hat{a}$. Since
$a$ grows over time whereas $\hat{a}$ remains relatively stable,
the dilaton field declines in the intergalactic space while it remains
largely unchanged within galaxies.\vskip4pt

For simplicity, we model the local scale $\hat{a}$ as homogeneous
within a galaxy, and allow it to merge with the global scale $a$
at the galaxy's boundary. Importantly, the local scale of the Milky
Way may differ from the local scale of the galaxy hosting a specific
SNeIa being observed. This is because a gravitationally bound galaxy
lives on an FLRW cosmic background that is expanding, rather than
static. As a result, its local scale $\hat{a}$ might, in principle,
experience modest growth in response to increases in the global scale
$a$. Therefore, it is reasonable to model the local scale $\hat{a}$
of a galaxy as a universal function (to be determined) of the redshift
$z$ of the galaxy, supplemented by a negligible idiosyncratic component.\vskip4pt

Consequently, as the dilaton field $\chi$ varies across the boundaries
of galaxies, the speed of light also varies at the boundaries due
to the relationship $c\propto\chi^{1/2}$. Figures \ref{fig:wavetrain-full}
depicts an intuitive schematic of a lightwave emitted from an SNeIa
as it propagates to the Earth-based observer. On its journey, the
lightwave undergoes 3 transits:\vskip4pt

{*} \emph{Transit \#1:} The lightwave emitted from an SNeIa residing
``inside host galaxy'', which is gravitationally bound and characterized
by a local scale $\hat{a}_{em}$, must first exit into the surrounding
intergalactic space, characterized by a global scale $a_{em}$.\vskip4pt

{*} \emph{Transit \#2:} The lightwave then traverses the null geodesic
of the FLRW metric and expands along with the cosmic scale factor
$a(t)$ of the intergalactic space until it reaches the outskirts
of the Milky Way, where the global scale is $a_{ob}$.\vskip4pt

{*} \emph{Transit \#3:} The lightwave enters the Milky Way, which
is gravitationally bound and characterized by a local scale $\hat{a}_{ob}$,
and finally reaches the Earth-based astronomer's telescope.\vskip4pt

While the middle stage of this journey, Transit \#2, is well understood
in standard cosmology, \emph{the first and last stages have been overlooked}
in previous VSL studies, seriously undermining their analyses and
conclusions. In the context of VSL, these stages are crucial due to
the additional refraction effects that occur at the boundaries of
the host galaxy and the Milky Way.\vskip8pt

\begin{figure}[t]
\noindent \begin{centering}
\hskip-6pt\includegraphics[scale=0.86]{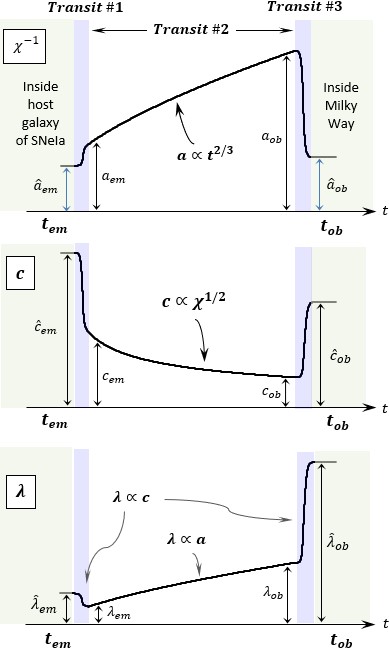}
\par\end{centering}
\caption{Variations of inverse dilaton (upper panel), speed of light (middle
panel), and wavelength (lower panel) along the lightwave trajectory
from emission to observation.}

\label{fig:variations}
\end{figure}

Figure \ref{fig:variations} illustrates the typical behavior of $\chi^{-1}$,
$c$, and the wavelength $\lambda$ along a lightwave trajectory.
In the top panel, it can be expected that $a_{em}>\hat{a}_{em}$ (since
the host galaxy resists cosmic expansion), $a_{ob}>a_{em}$ (due to
the expansion of intergalactic space), and $\hat{a}_{ob}<a_{ob}$
(since the Milky Way also resists cosmic expansion). Quantitatively,
we can deduce the variation of wavelength during the three transits
as follows:\vskip4pt

\begin{itemize}
\item \emph{The emission event:} an SNeIa radiates a wavetrain with a specific
wavelength $\hat{\lambda}_{em}.$\vskip4pt
\item \emph{Transit \#1:} The wavetrain exits the host galaxy to enter the
surrounding intergalactic space. During this transition, its wavelength
is compressed to $\lambda_{em}$ due to the reduction in the speed
of light from $\hat{c}_{em}$ to $c_{em}$ across the host galaxy's
boundary, viz.
\begin{equation}
\frac{\lambda_{em}}{\hat{\lambda}_{em}}=\frac{c_{em}}{\hat{c}_{em}}=\frac{a_{em}^{-1/2}}{\hat{a}_{em}^{-1/2}}\label{eq:transit1}
\end{equation}
Appendix \ref{sec:Refraction-effect} summarizes the components involved
in the refraction that is induced by variations in the velocity of
wavetrains.
\item \emph{Transit \#2:} The wavetrain follows the null geodesics of the
FLRW metric. As it approaches the outskirts of the Milky Way, its
wavelength has expanded from $\lambda_{em}$ to $\lambda_{ob}$, as
given in Eq. \eqref{eq:transit2}, viz.
\[
\hskip95pt\frac{\lambda_{ob}}{\lambda_{em}}=\frac{a_{ob}}{a_{em}}\tag{see Eq. \eqref{eq:transit2}}
\]
\item \emph{Transit \#3:} The wavetrain enters the Milky Way and reaches
the astronomer's telescope. Its wavelength is further prolonged due
to an increase in the speed of light from $c_{ob}$ to $\hat{c}_{ob}$
across the Milky Way's boundary, viz.
\begin{equation}
\frac{\hat{\lambda}_{ob}}{\lambda_{ob}}=\frac{\hat{c}_{ob}}{c_{ob}}=\frac{\hat{a}_{ob}^{-1/2}}{a_{ob}^{-1/2}}\,.\label{eq:transit3}
\end{equation}
\end{itemize}

\subsection{\label{subsec:Benchmark}The ``benchmark'' wavelength}

There is one more crucial element to consider. In calculating the
redshift of an SNeIa, it would be \emph{incorrect} to directly compare
the observed wavelength $\hat{\lambda}_{ob}$ with the emitted wavelength
$\hat{\lambda}_{em}$. This is because $\hat{\lambda}_{em}$ is associated
with the emission event occurring inside the host galaxy, and the
observer cannot directly measure $\hat{\lambda}_{em}$ since she is
located within the Milky Way. If the SNeIa were situated inside the
Milky Way, it would emit a wavelength $\lambda^{*}$ that differs
from $\hat{\lambda}_{em}$, as the two galaxies can have different
values of local scales, $\hat{a}_{em}$ versus $\hat{a}_{ob}$. The
wavelength $\lambda^{*}$, which the observer \emph{can} measure,
is the ``benchmark'' wavelength to be compared with the observed $\hat{\lambda}_{ob}$
in calculating the redshift.\vskip4pt

To illustrate this issue, let us recall that the lengthscale of any
physical process is inversely proportional to the dilaton field, according
to Eq. \eqref{eq:length-vs-chi} in Section \ref{sec:Generating-VSL}.
For a galaxy, the dilaton field is in turn inversely proportional
to the local scale of that galaxy. Consider two identical atoms, one
located inside the host galaxy and the other within the Milky Way.
If the atom in the host galaxy emits a lightwave with wavelength $\hat{\lambda}_{em}$,
its counterpart in the Milky Way emits an \emph{identical} lightwave
but with wavelength $\lambda^{*}$ \emph{adjusted} to the Milky Way's
local scale. The following equality holds
\begin{equation}
\frac{\lambda^{*}}{\hat{\lambda}_{em}}=\frac{\hat{a}_{ob}}{\hat{a}_{em}}\label{eq:def-yardstick}
\end{equation}

As shown in the right side of Figure \ref{fig:wavetrain-full}, the
observer must compare $\hat{\lambda}_{ob}$ with her ``benchmark''
wavelength $\lambda^{*}$. Finally, the observer calculates the redshift
$z$ as the relative change between the observed wavelength and the
``benchmark'' wavelength, given by
\begin{equation}
z:=\frac{\hat{\lambda}_{ob}-\lambda^{*}}{\lambda^{*}}\,.\label{eq:def-redshift}
\end{equation}

We should note that allowing $\hat{a}_{ob}$ to differ from $\hat{a}_{em}$
creates a potential pathway to resolving the $H_{0}$ tension, a topic
that will be discussed in Section \ref{sec:Resolving-H0-tension}.

\section{\label{sec:Modifications-all-formulae}Modifying redshift formulae
and Hubble law using Varying $\large{\textbf{c}}$}

We are now fully equipped to derive cosmographic formulae applicable
to our VSL cosmology.

\subsection{\label{subsec:Modified-Lemaitre}Modifying the Lema\^itre redshift
formula}

What is remarkable in the demonstration depicted in Figure \ref{fig:wavetrain-full}
is that the stretching of the wavecrest during Transit \#3 does \emph{not}
cancel out the compression of the wavecrest during Transit \#1. The
net effect of the two transits increases the value of $z$ and results
in a new formula for the redshift. Below is our derivation.\vskip4pt

Combining Eq. \eqref{eq:def-redshift} with Eqs. \eqref{eq:transit1},
\eqref{eq:transit2}, and \eqref{eq:transit3}, we obtain
\begin{align}
1+z & =\frac{\hat{\lambda}_{ob}}{\lambda^{*}}=\frac{\hat{\lambda}_{ob}}{\lambda_{ob}}.\frac{\lambda_{ob}}{\lambda_{em}}.\frac{\lambda_{em}}{\hat{\lambda}_{em}}.\frac{\hat{\lambda}_{em}}{\lambda^{*}}=\frac{a_{ob}^{3/2}}{a_{em}^{3/2}}.\frac{\hat{a}_{em}^{3/2}}{\hat{a}_{ob}^{3/2}}
\end{align}

Defining the ratio of local scales as a function of redshift:
\begin{equation}
\frac{\hat{a}_{em}}{\hat{a}_{ob}}:=F(z)\label{eq:def-F}
\end{equation}
where $F(z=0)=1$, and setting
\begin{equation}
a:=\frac{a_{em}}{a_{ob}}
\end{equation}
we arrive at the \emph{modified Lema\^itre redshift formula}: 
\begin{equation}
1+z=a^{-3/2}\,F^{3/2}(z)\label{eq:modified-Lemaitre}
\end{equation}
If $F(z)\equiv1\ \forall z$, viz. all galaxies have the same local
scale, the modified Lema\^itre redshift formula simplifies to:
\begin{equation}
1+z=a^{-3/2}\label{eq:modified-Lemaitre-simplified}
\end{equation}
These formulae are decisively different from the classic Lema\^itre
redshift formula, $1+z=a^{-1}$. The $3/2$--exponent in the modified
Lema\^itre formulae arises as a result of the anisotropic time scaling
in Eq. \eqref{eq:tau-vs-length}.\vskip4pt

It is essential to emphasize that the alteration in wavelength---due
to the refraction effect across boundaries of galaxies---is instrumental
in enabling the VSL effects to manifest in the modified Lema\^itre
redshift formula. To the best of our knowledge, existing VSL analyses
in the literature have not considered this wavelength alteration.
This omission hinders theirs ability to detect the effects of VSL
on the Hubble diagram of SNeIa and late-time cosmic acceleration.

\subsection{\label{subsec:Modified-Hubble}Modifying the Hubble law: An emergent
multiplicative factor of $3/2$}

The current-time Hubble constant $H_{0}$ is defined as
\begin{equation}
H_{0}:=\frac{1}{a}\frac{da}{dt}|_{t=t_{0}}
\end{equation}
For a low-$z$ emission source, this yields
\begin{equation}
a=1+H_{0}(t-t_{0})+\dots
\end{equation}

\noindent Let $d=c_{0}.(t_{0}-t)$ represent the distance from Earth
to the emission source, and note that $F(z)\simeq1$ for low $z$.
For small $z$ and $d$, the Taylor expansion for the modified Lema\^i
redshift formula obtained in Eq. \eqref{eq:modified-Lemaitre} produces
the \emph{modified Hubble law}: 
\begin{equation}
z=\frac{3}{2}H_{0}\,\frac{d}{c_{0}}\label{eq:modified-hubble-law}
\end{equation}
In comparison to the classic Hubble law, where the speed of light
is explicitly restored: 
\begin{equation}
z_{\,\text{(classic)}}=H_{0}\,\frac{d}{c}\label{eq:hubble-law}
\end{equation}
the modified Hubble law \emph{acquires a multiplicative prefactor
of 3/2.} A significant consequence of this adjustment is a (re)-evaluation
of the Hubble constant $H_{0}$, which has implications for BDRS's
CMB analysis and the age problem---topics that will be discussed
in Sections \ref{sec:BDRS} and \ref{sec:Resolve-age-prob}.

\subsection{\label{subsec:Modified-d-z}Modifying the distance--redshift formula}

Using the evolution $a\propto t^{2/3}$ per Eq. \eqref{eq:a-vs-t},
we can derive that
\begin{equation}
H(t)=\frac{1}{a}\frac{da}{dt}=\frac{2}{3\,t}=H_{0}a^{-3/2}\label{eq:here1}
\end{equation}
The modified Lema\^itre redshift formula, Eq. \eqref{eq:modified-Lemaitre},
can be recast as
\begin{equation}
\ln\frac{(1+z)^{2/3}}{F}=-\ln a\label{eq:dz}
\end{equation}
and, with the aid of Eq. \eqref{eq:here1}, renders 
\begin{equation}
d\ln\frac{(1+z)^{2/3}}{F}=-H_{0}a^{-3/2}dt\label{eq:tmp}
\end{equation}
For the modified FLRW metric described in Eq. \eqref{eq:modified-RW},
the coordinate distance in flat space is
\begin{equation}
r=c_{0}\int_{t_{em}}^{t_{ob}}\frac{dt}{a^{3/2}(t)}\label{eq:coordinate-distance}
\end{equation}
From Eqs. \eqref{eq:tmp} and \eqref{eq:coordinate-distance}, and
noting that $F(z=0)=1$, we obtain the \emph{modified} distance--redshift
formula in a compact expression
\begin{align}
\frac{r}{c_{0}} & =\frac{1}{H_{0}}\ln\frac{(1+z)^{2/3}}{F}\,.\label{eq:modified-r-vs-z}
\end{align}

\subsection{\label{subsec:Modified-dL-z}Modifying the luminosity distance--redshift
formula}

In standard cosmology, the luminosity distance $d_{L}$ is defined
via the absolute luminosity $L$ and the apparent luminosity $J$:
\begin{equation}
d_{L}^{2}=\frac{L}{4\pi J}\label{eq:definition-of-dL}
\end{equation}
The absolute luminosity $L$ and the apparent luminosity $J$ are
related as
\begin{equation}
4\pi r^{2}J=L\frac{\hat{\lambda}_{em}}{\hat{\lambda}_{ob}}.\frac{\hat{\lambda}_{em}}{\hat{\lambda}_{ob}}\label{eq:J-vs-L}
\end{equation}
In the right hand side of Eq. \eqref{eq:J-vs-L}, the first term $\hat{\lambda}_{em}/\hat{\lambda}_{ob}$
represents the ``loss'' in energy of the redshifted photon known as
the ``Doppler theft'' \footnote{Note: This energy loss is consistent with the scaling property of
energy for our VSL mechanism, as described by Eq. \eqref{eq:length-vs-chi}
in Section \ref{sec:Generating-VSL}. In intergalactic space, the
energy of the traveling photon scales as $E\propto\chi\propto a^{-1}$,
leading to a decline in energy as the universe expands.}. The second (identical) term $\hat{\lambda}_{em}/\hat{\lambda}_{ob}$
arises from the dilution factor in photon density, as the same number
of photons is distributed over a prolonged wavecrest in the radial
direction (i.e., along the light ray). The $4\pi r^{2}$ in the left
hand side of Eq. \eqref{eq:J-vs-L} accounts for the spherical dilution
in flat space. From \eqref{eq:definition-of-dL} and \eqref{eq:J-vs-L},
we obtain
\begin{equation}
d_{L}=r\,\frac{\hat{\lambda}_{ob}}{\hat{\lambda}_{em}}
\end{equation}
Using the definitions of redshift and the ``benchmark'' wavelength,
Eqs. \eqref{eq:def-redshift} and \eqref{eq:def-yardstick} respectively,
the luminosity distance becomes
\begin{equation}
d_{L}=r\,\frac{\hat{\lambda}_{ob}}{\lambda^{\star}}.\frac{\lambda^{\star}}{\hat{\lambda}_{em}}=r\,(1+z)\,\frac{\hat{a}_{ob}}{\hat{a}_{em}}
\end{equation}
or, by including \eqref{eq:def-F}:
\begin{equation}
d_{L}=r\,(1+z)\,\frac{1}{F(z)}\label{eq:tmp-3}
\end{equation}

\noindent Due to the refraction effect during Transit \#3, the apparent
luminosity distance observed by the Earth-based astronomer $\hat{d}_{L}$
differs from $d_{L}$ by the factor $\hat{c}_{ob}/c_{ob}$, viz.

\begin{align}
\frac{\hat{d}_{L}}{\hat{c}_{ob}} & =\frac{d_{L}}{c_{ob}}\label{eq:here2}
\end{align}
Finally, combining Eqs. \eqref{eq:modified-r-vs-z}, \eqref{eq:tmp-3},
and \eqref{eq:here2}, we arrive at the \emph{modified }luminosity
distance--redshift relation:
\begin{equation}
\frac{\hat{d}_{L}}{\hat{c}_{ob}}=\frac{1+z}{H_{0}F(z)}\,\ln\frac{(1+z)^{2/3}}{F(z)}\label{eq:modified-dL-vs-z}
\end{equation}
where $\hat{d}_{L}$ is the luminosity distance observed by the Earth-based
astronomer and $\hat{c}_{ob}$ the speed of light measured in the
Milky Way (i.e., $300,000$ km/s). Formula \eqref{eq:modified-dL-vs-z}
contains a single parameters $H_{0}$ and involves a function $F(z)$
that captures the evolution of the local scale of galaxies as a function
of redshift.

\section{\label{sec:Reanalysis-Pantheon}Re-analyzing $\,$Pantheon Catalog
using Varying $\large{\textbf{c}}$}

This section applies the new formula, Eq. \eqref{eq:modified-dL-vs-z},
to the Combined Pantheon Sample of SNeIa. In \citep{Scolnic-2018},
Scolnic and collaborators produced a dataset of apparent magnitudes
for $1,048$ SNeIa with redshift $z$ ranging from $0.01$ to $2.25$,
accessible in \citep{Pantheon-data}. For each SNeIa $i^{th}$, the
catalog provides the redshift $z_{i}$, the apparent magnitude $m_{B,i}^{\text{Pantheon}}$
together with its error bar $\sigma_{i}^{\text{Pantheon}}$. We apply
the absolute magnitude $M=-19.35$ to compute the distance modulus,
$\mu^{\text{Pantheon}}:=m_{B}^{\text{Pantheon}}-M$. The distance
modulus is then converted to the luminosity distance $d_{L}$ using
the following relation:
\begin{equation}
\mu=5\log_{10}(d_{L}/\text{Mpc})+25\label{eq:modulus}
\end{equation}
The Pantheon data, along with their error bars, are displayed in the
Hubble diagram shown in Fig. \ref{fig:fit-without-F}.

\subsection{$\Lambda$CDM and standard EdS as benchmarking models}

For benchmarking purposes, we first fit the Pantheon Catalog with
the flat $\Lambda$CDM model. The luminosity distance--redshift relation
for this model is a well-established result (where $\Omega_{M}+\Omega_{\Lambda}=1$)
\begin{equation}
\frac{d_{L}}{c}=\frac{1+z}{H_{0}}\,\int_{0}^{z}\frac{dz'}{\sqrt{\Omega_{M}(1+z')^{3}+\Omega_{\Lambda}}}\label{eq:LCDM-formula}
\end{equation}
Our fit will minimize the normalized error
\begin{equation}
\chi^{2}:=\frac{1}{N}\sum_{j=1}^{N}\Biggl(\frac{\mu_{j}^{\text{model}}-\mu_{j}^{\text{Pantheon}}}{\sigma_{j}^{\text{Pantheon}}}\Biggr)^{2}\label{eq:chi-1}
\end{equation}
with the sum taken over all $N=1,048$ Pantheon data points. The best
fit for the $\Lambda$CDM model yields $H_{0}=70.2$ km/s/Mpc, $\Omega_{M}=0.285$,
$\Omega_{\Lambda}=0.715$, with the minimum error $\chi_{\text{min}}^{2}(\Lambda\text{CDM})=0.98824$.
The $d_{L}$--$z$ curve for the $\Lambda$CDM model is depicted
by the dashed line in Fig. \ref{fig:fit-without-F}.\vskip4pt

Also for benchmarking purposes, we consider a ``fiducial'' model:
the standard EdS universe (i.e. with \emph{constant} speed of light).
The luminosity distance--redshift formula for this fiducial model
can be obtained by setting $\Omega_{\Lambda}=0$ and $\Omega_{M}=1$
in Eq. \eqref{eq:LCDM-formula}, yielding
\begin{equation}
\frac{d_{L}}{c}=2\frac{1+z}{H_{0}}\,\Bigl(1-\frac{1}{\sqrt{1+z}}\Bigr)\label{eq:EdS-formula}
\end{equation}
Figure \ref{fig:fit-without-F} displays the $d_{l}$--$z$ curve
as a dotted line for the fiducial EdS model (using the $H_{0}=70.2$
value obtained above for the $\Lambda$CDM model). This curve fits
well with the Pantheon data for low $z$ but fails to capture the
data for high $z$. The Pantheon data with $z\gtrsim0.1$ show an
excess in the distance modulus compared with the baseline EdS model,
meaning that high--redshift SNeIa appear dimmer than predicted by
the fiducial EdS model. As a result, this discrepancy necessitated
the introduction of the $\Lambda$ component, commonly referred to
as dark energy, characterized by an equation of state\linebreak $w=-1$
and an energy density of $\Omega_{\Lambda}\approx0.7$.

\subsection{\label{subsec:Disabling-F}Fitting with VSL model: Disabling $F(z)$}

\begin{figure}[!t]
\begin{centering}
\hskip-5pt\includegraphics[scale=0.75]{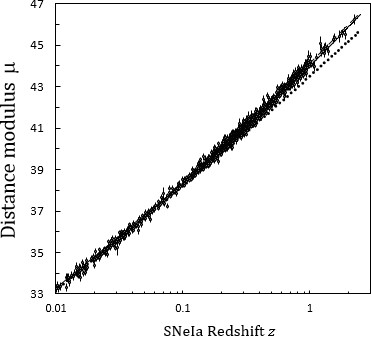}
\par\end{centering}
\caption{Fitting Pantheon using Formula \eqref{eq:modified-dL-vs-z-w/o-F}.
Open circles: 1,048 data points with error bars. Solid line: our Formula
\eqref{eq:modified-dL-vs-z-w/o-F} with $H_{0}=44.4$. Dashed line:
$\Lambda$CDM Formula \eqref{eq:LCDM-formula} with $H_{0}=70.2$,
$\Omega_{\Lambda}=0.715$. Dotted line: EdS Formula \eqref{eq:EdS-formula}
with $H_{0}=70.2$.}

\label{fig:fit-without-F}
\end{figure}

In this subsection, we will disable the evolution of the local scale
of galaxies by setting $F(z)\equiv1$ in Formula \eqref{eq:modified-dL-vs-z}.
This means that the fit is carried out with respect to a simplified
formula with one adjustable parameter $H_{0}$:\linebreak
\begin{equation}
\frac{\hat{d}_{L}}{\hat{c}_{ob}}=\frac{1+z}{\frac{3}{2}H_{0}}\,\ln(1+z)\label{eq:modified-dL-vs-z-w/o-F}
\end{equation}
Hereafter, the luminosity distance $\hat{d}_{L}$ observed by the
Earth-based astronomer will be used in the conversion described by
Eq. \eqref{eq:modulus}.\vskip4pt

The best fit of the Pantheon data to this formula yields $H_{0}=44.4\text{ km/s/Mpc}$,
corresponding to $\chi_{\text{min}}^{2}=1.25366$. Figure \ref{fig:fit-without-F}
displays our fit as the solid line. Although this fit performs worse
than the $\Lambda$CDM model, which has $\chi_{\text{min}}^{2}(\Lambda\text{CDM})=0.98824$,
it substantially reduces the excess in distance moduli for $z\gtrsim0.1$
compared with the ``fiducial'' EdS model, as shown in Fig. \ref{fig:fit-without-F}.\vskip4pt

We must emphasize that both Formulae \eqref{eq:EdS-formula} and \eqref{eq:modified-dL-vs-z-w/o-F}
are one-parameter models. Both models are based on an EdS universe,
but our VSL model accommodates varying speed of light, whereas the
``fiducial'' EdS model operates under the assumption of a constant
speed of light. Therefore, we can conclude that \emph{varying speed
of light is responsible for the improved performance of our VSL model}
compared to the ``fiducial'' EdS model.\vskip4pt

This aspect can be explained as follows. In the high $z$ limit, Formula
\eqref{eq:EdS-formula} of the ``fiducial'' EdS model yields
\begin{equation}
d_{L}\simeq z\label{eq:b1}
\end{equation}
whereas Formula \eqref{eq:modified-dL-vs-z-w/o-F} of our VSL model
gives
\begin{equation}
d_{L}\propto\hat{d}_{L}\simeq z\,\ln z\label{eq:b2}
\end{equation}
The additional $\ln z$ term in Eq. \eqref{eq:b2} compared to Eq.
\eqref{eq:b1} induces a steeper slope in the high--$z$ portion
of the $d_{L}$--$z$ curve, which translates to an excess in distance
modulus at high redshift. Notably, our VSL model does not require
dark energy whatsoever to account for this behavior.\vskip4pt

The performance of our VSL model can be improved by enabling the function
$F(z)$, which involves allowing the local scales of galaxies to evolve.
This task will be carried out in the following subsections.\pagebreak

\subsection{\label{subsec:Enabling-F-bin}Enabling $F(z)$: $\ $A binning approach}

\noindent 
\begin{figure}[!t]
\begin{centering}
\includegraphics[scale=0.75]{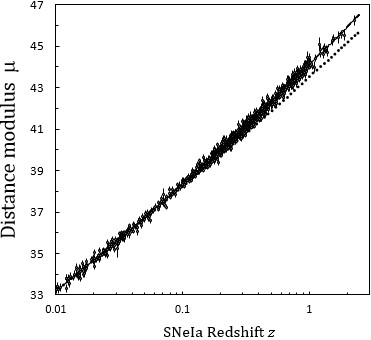}
\par\end{centering}
\caption{Fitting Pantheon using Formula \eqref{eq:eff-F-sim}. Open circles:
1,048 data points with error bars. Solid line: our Formula \eqref{eq:eff-F-sim}
with $H_{0}=46.6$ and the $F^{(i)}$ values given in Table \ref{table:value-eff-F}.
Dashed line: $\Lambda$CDM Formula \eqref{eq:LCDM-formula} with $H_{0}=70.2$,
$\Omega_{\Lambda}=0.715$. Dotted line: EdS Formula \eqref{eq:EdS-formula}
with $H_{0}=70.2$.}

\label{fig:fit-w-F-binning}
\end{figure}

\noindent 
\begin{table}[t]
\noindent \begin{centering}
\begin{tabular}{|ccccc|}
\hline 
\multicolumn{1}{|c|}{$\ $Bin$\ $} & \multicolumn{1}{c|}{$\ $Number of SNeIa$\ $} & \multicolumn{1}{c|}{$\ $Min $z$$\ $} & \multicolumn{1}{c|}{$\ $Max $z$$\ $} & $\ $$F^{(i)}$$\ $\tabularnewline
\hline 
\#1 & 105 & 0.010 & 0.032 & 1.000\tabularnewline
\#2 & 105 & 0.032 & 0.096 & 1.000\tabularnewline
\#3 & 105 & 0.099 & 0.159 & 0.997\tabularnewline
\#4 & 105 & 0.159 & 0.203 & 0.996\tabularnewline
\#5 & 105 & 0.203 & 0.249 & 0.994\tabularnewline
\#6 & 105 & 0.249 & 0.299 & 0.994\tabularnewline
\#7 & 105 & 0.300 & 0.366 & 0.988\tabularnewline
\#8 & 105 & 0.368 & 0.508 & 0.984\tabularnewline
\#9 & 105 & 0.510 & 0.742 & 0.978\tabularnewline
\#10 & 103 & 0.743 & 2.260 & 0.968\tabularnewline
\hline 
\end{tabular}
\par\end{centering}
\caption{Values of $F^{(i)}$ for the 10 respective bins.}
\label{table:value-eff-F}
\end{table}
\begin{figure}[t]
\begin{centering}
\hskip-6pt\includegraphics[scale=0.75]{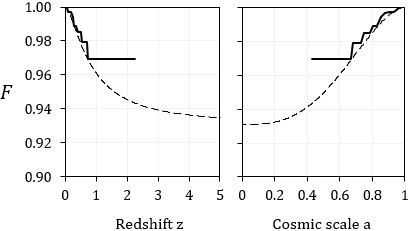}
\par\end{centering}
\caption{The variation of $F(z)$ as functions of redshift (left panel) and
cosmic scale factor (right panel). Solid staircase lines are the result
obtained in Section \ref{subsec:Enabling-F-bin}. Dashed lines are
the result obtained in Section \ref{subsec:Enabling-F-analytic}.}

\label{fig:F-vs-z-and-a}
\end{figure}

\vspace{-1cm}In this subsection, we will incorporate the variation
in the local scales of galaxies, as characterized by $F(z)$. Developing
model for $F(z)$ would require knowledge of galactic formation and
structures. Here, we avoid that complexity by extracting $F(z)$ directly
from the Pantheon data. To do so, we spit the Pantheon dataset into
10 bins ordered by increasing redshift. Bins $\#1$ to $\#9$ each
contains 105 data points, while Bin $\#10$ contains 103 data points,
totaling 1,048 data points. The range of redshift for each bin is
given in Table \ref{table:value-eff-F}.\vskip4pt

All Pantheon data points in Bin $\#i$ are treated as having a common
value of $F^{(i)}$ for the function $F(z)$. For a data point $\#j$
that belongs to Bin $\#i$, Formula \eqref{eq:modified-dL-vs-z} reads
\begin{equation}
\frac{\hat{d}_{L,j}}{\hat{c}_{ob}}=\frac{1+z_{j}}{H_{0}F^{(i)}}\,\ln\frac{(1+z_{j})^{2/3}}{F^{(i)}}\label{eq:eff-F-sim}
\end{equation}

Instead of fitting each bin separately, we impose one common value
for $H_{0}$ across all bins. The fit thus involves $H_{0}$ and 10
values for $\{F^{(i)},\,i=1..10\}$. Figure \ref{fig:fit-w-F-binning}
displays our fit to Formula \eqref{eq:eff-F-sim}. The best fit yields
$H_{0}=46.6\text{ km/s/Mpc}$, and the values of $F^{(i)}$ are given
in the last column of Table \ref{table:value-eff-F}. The minimum
error is $\chi_{\text{min}}^{2}=0.97803$. The staircase lines in
Fig. \ref{fig:F-vs-z-and-a} depict $F^{(i)}$ as function of $z$
and the cosmic factor $a$.\linebreak

The values $F^{(i)}$ reveal a monotonic decrease with respect to
redshift, or equivalently, a monotonic increase in terms of $a$.
This behavior indicates that the local scales of galaxies gradually
grow during the course of cosmic expansion, implying that galaxies
cannot fully resist this expansion. From $z\simeq2$ to the present
time, galaxies have slightly expanded by about 3\%, during which process
the cosmic scale factor has approximately doubled, i.e. $a|_{z\simeq2}\simeq0.5$
with $F(z\simeq2)\approx0.969$.

\subsection{\label{subsec:Enabling-F-analytic}Enabling $F(z)$: $\ $A parametrization
approach}

\noindent 
\begin{figure}[!t]
\begin{centering}
\includegraphics[scale=0.75]{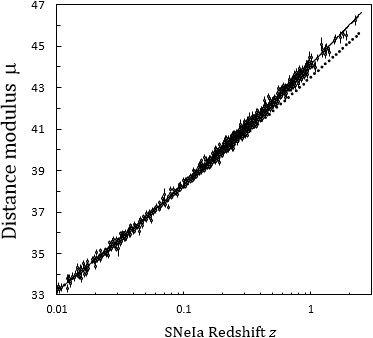}
\par\end{centering}
\caption{Fitting Pantheon using Formula \eqref{eq:modified-dL-vs-z}. Open
circles: 1,048 data points with error bars. Solid line: our Formula
\eqref{eq:modified-dL-vs-z} with $H_{0}=47.22$, and $F(z)$ given
in Eq. \eqref{eq:func-form-of-F} with $b_{1}=b_{2}=2$ and $F_{\infty}=0.931$.
Dashed line: $\Lambda$CDM Formula \eqref{eq:LCDM-formula} with $H_{0}=70.2$,
$\Omega_{\Lambda}=0.715$. Dotted line: EdS Formula \eqref{eq:EdS-formula}
with $H_{0}=70.2$.}

\label{fig:fit-with-F}
\end{figure}

The steady decline of $F^{(i)}$ across the 10 bins with respect to
redshift suggests adopting the following parametrization for $F(z)$:
\begin{equation}
F(z)=1-(1-F_{\infty})\left(1-(1+z)^{-b_{1}}\right)^{b_{2}}\label{eq:func-form-of-F}
\end{equation}
with $b_{1}\in\mathbb{R}^{+}$, $b_{2}\in\mathbb{R}^{+}$, and $F_{\infty}\in[0,1]$,
supporting a monotonic interpolation from $F(z=0)=1$ to $F(z\rightarrow\infty)=F_{\infty}$.
After some experimentation, we find that setting $b_{1}=b_{2}=2$
offers good overall performance.\vskip4pt

We will apply Formula \eqref{eq:modified-dL-vs-z} in conjunction
with \eqref{eq:func-form-of-F} (with\textbf{ $b_{1}=b_{2}=2$}) to
fit the Pantheon dataset. The best fit is displayed in Fig. \ref{fig:fit-with-F},
yielding $H_{0}=47.22$ and $F_{\infty}=0.931$. The minimum error
is $\chi_{\text{min}}^{2}=0.98556$, a performance that is competitive
with---if not exceeding---that of the $\Lambda$CDM model, which
has $\chi_{\text{min}}^{2}(\Lambda\text{CDM})=0.98824$.\vskip4pt

\begin{figure}[!t]
\begin{centering}
\hskip-6pt\includegraphics[scale=0.55]{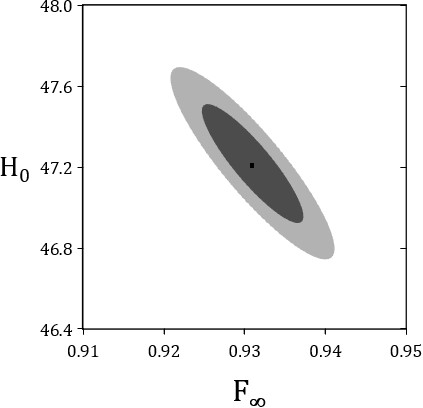}
\par\end{centering}
\caption{Joint distribution of $H_{0}$ and $F_{\infty}$, showing $68\%$
CL and 95\% CL regions. Peak occurs at $H_{0}=47.22,\ F_{\infty}=0.931$.}

\noindent \label{fig:jointdist-H0-Finfty}
\end{figure}

Figure \ref{fig:F-vs-z-and-a} shows the variation of $F$ in dashed
lines with respect to both redshift and the cosmic factor. We also
produce the joint distribution for $H_{0}$ and $F_{\infty}$, as
shown in Figure \ref{fig:jointdist-H0-Finfty}, yielding $H_{0}=47.21\pm0.4\text{ km/s/Mpc}$
(95\% CL) and $F_{\infty}=0.931\pm0.008$ (95\% CL). This value of
$F_{\infty}$ indicates that the local scales of galaxies have increased
by approximately 7\% since the formation of the first stable galaxies
(i.e., those at the largest redshift).\vskip16pt
\begin{center}
\textbf{\emph{Comparison of VSL approach with $\Lambda$CDM model\vskip12pt}}
\par\end{center}

Our VSL fit, in effect, involves two parameters: $H_{0}$ and $F_{\infty}$---the
same number of parameters as the $\Lambda$CDM model ($H_{0}$ and
$\Omega_{\Lambda}$). However, the parameter $F_{\infty}$ has a well-defined
\emph{astrophysical} meaning; it denotes the local scales of the first
stable galaxies in comparison to the local scale of the Milky Way.
Moreover, the function $F(z)$, which captures the evolution of the
local scales of galaxies during cosmic expansion, plays a role in
a potential resolution of the $H_{0}$ tension---a topic that will
be discussed in Section \ref{sec:Resolving-H0-tension}.\vskip4pt

In contrast, the $\Lambda$CDM model requires a $\Lambda$ component,
the nature of which is still not understood. Its energy density value
$\Omega_{\Lambda}\approx0.7$ also raises a coincidence problem. Furthermore,
the $\Lambda$CDM model currently encounters the $H_{0}$ tension.
If the $\Lambda$ component is treated as dynamical---an approach
explored in several ongoing efforts to resolve the $H_{0}$ tension---this
would introduce an array of new parameters to the $\Lambda$CDM model.

\subsection{\label{subsec:H0-reduction}The cause for the reduction in $H_{0}$
value}

In the $z\rightarrow0$ limit, Eq. \eqref{eq:func-form-of-F} with
$b_{1}=b_{2}=2$ can be approximated as
\begin{equation}
F(z)\simeq1-4(1-F_{\infty})\,z^{2}+\dots
\end{equation}
From this, Formula \eqref{eq:modified-dL-vs-z} then gives
\begin{equation}
\frac{\hat{d}_{L}}{\hat{c}_{ob}}=\frac{2}{3H_{0}}\,z+\mathcal{O}(z^{2})
\end{equation}
leading to the \emph{modified Hubble law}: 
\begin{equation}
z=\frac{3}{2}H_{0}\,\frac{\hat{d}_{L}}{\hat{c}_{ob}}\label{eq:modified-hubble-law-3}
\end{equation}
This result aligns with Eq. \eqref{eq:modified-hubble-law} derived
in Section \ref{subsec:Modified-Hubble} based on the modified Lema\^itre
redshift formula, Eq. \eqref{eq:modified-Lemaitre}. In contrast to
the classic Hubble law, where the speed of light is constant: 
\[
z_{\,\text{(classic)}}=H_{0}\,\frac{d_{L}}{c}\tag{see Eq. \eqref{eq:hubble-law}}
\]
the modified Hubble law \emph{acquires a multiplicative factor of
3/2.} Hence, in our VSL cosmology, low-redshift emission sources exhibit
a linear relationship between $z$ and the luminosity distance,\emph{
but characterized by a coefficient of $\frac{3}{2}H_{0}$ rather than
$H_{0}$}.\emph{\vskip4pt}

Since a linear-line fit of $z$ on $d_{L}$ for low-redshift emission
sources is known to yield a slope of approximately $70$, the resulting
value of $H_{0}$ obtained through our VSL approach is thus only $2/3$
of this value, specifically $H_{0}\approx47$, rather than $H_{0}\approx70$
as predicted by standard cosmology.

\section{\label{sec:VSL-as-alternative}A new interpretation: $\,$Variable
speed of light as an alternative to dark energy and cosmic acceleration}

The Hubble diagram of SNeIa has been interpreted as a definitive hallmark
of late-time accelerated expansion, providing the (only) \emph{direct}
evidence for dark energy. These stellar explosions serve as ``standard
candles'' due to their consistent peak brightness, allowing astronomers
to determine distances to the galaxies in which they reside. In the
late 1990s, two independent teams, the High-Z Supernova Search Team
\citep{Riess-1998} and the Supernova Cosmology Project \citep{Perlmutter-1999},
measured the apparent brightness of distant SNeIa, finding them dimmer
than expected based on the EdS model, which describes a flat, expanding
universe dominated by matter. This can be seen in the Hubble diagram
of SNeIa (see Fig. \eqref{fig:fit-w-F-binning}), in which the section
with $z\gtrsim0.1$ exhibits an distance modulus greater than that
predicted by the EdS model. This behavior has been interpreted as
indicating that the expansion of the universe is accelerating rather
than decelerating.\vskip4pt

However, our quantitative analysis of SNeIa in the preceding section
offers a new interpretation as an \emph{alternative} to late-time
acceleration. Mathematically, as explained in Section \ref{subsec:Disabling-F},
in the high $z$ limit, the EdS universe yields 
\[
\hskip95ptd_{L}\simeq z\tag{see Eq. \eqref{eq:b1}}
\]
whereas our VSL-based formula renders 
\[
\hskip95ptd_{L}\simeq z\,\ln z\tag{see Eq. \eqref{eq:b2}}
\]
Thus, high-redshift SNeIa acquire an additional factor of $\ln z$
compared with the EdS model. This results in an further upward slope
relative to that of the EdS model, successfully capturing the behavior
of SNeIa in the high-z section.\vskip8pt

\textbf{\emph{Physical intuition:}} There is a fundamental reason---based
on our VSL framework---behind this excess in distance modulus that
we will explain below. Consider two supernovae A and B at distances
$d_{A}$ and $d_{B}$ away from the Earth, such that $d_{B}=2\,d_{A}$.
In standard cosmology, their redshift values $z_{A}$ and $z_{B}$
are related by $z_{A}\approx2\,z_{A}$ (to first-order approximation).
However, this relation breaks down in the VSL context. In a VSL cosmology
which accommodates variation in the speed of light in the form�$c\propto a^{-1/2}$,
light traveled faster in the distant past (when the cosmic factor
$a\ll1$) than in the more recent epoch (when $a\lesssim1$). Therefore,
the photon emitted from supernova B was able to cover twice the distance
in \emph{less than} twice the time required for the photon emitted
from supernova A. Having spent less time in transit than what standard
cosmology would require, the B-photon experienced less cosmic expansion
than expected and thus a lower redshift than what the classic Lema\^itre
formula would dictate. Namely:
\begin{equation}
z_{B}<2\,z_{A}\ \ \ \text{for}\ \ d_{B}=2\,d_{A}
\end{equation}
Conversely, consider a supernova C with $z_{C}=2\,z_{A}$. For the
C-photon to experience twice the redshift of the A-photon, it must
travel a distance \emph{greater than} twice that of the A-photon,
viz.: $d_{C}>2\,d_{A}$. This is because since the C-photon traveled
faster at the beginning of its journey toward Earth, it must originate
from a farther distance (thus appearing fainter than expected) to
experience enough cosmic expansion and therefore the requisite amount
of redshift. Namely:
\begin{equation}
d_{C}>2\,d_{A}\ \ \ \text{for}\ \ ~z_{C}=2\,z_{A}
\end{equation}
Consequently, the SNeIa data exhibit an additional upward slope in
their Hubble diagram in the high-$z$ section.\vskip12pt

\textbf{\emph{Conclusion:}} Hence, \emph{a declining speed of light
presents a viable alternative to cosmic acceleration}, eliminating
the need for the $\Lambda$ component and dissolving its fine-tuning
and coincidence problems. In our VSL framework, during the universe
expansion, the dilaton field $\chi$ in intergalactic space decreases
and leads to a decline in $c$ (per $c\propto\chi^{1/2}\propto a^{-1/2}$),
affecting the propagation of lightwaves from distant SNeIa to the
observer. While the impact of a declining $c$ on lightwaves is negligible
within the Solar System and on galactic scales, it accumulates on
the cosmic scale and makes high-redshift SNeIa appear dimmer than
predicted by the standard EdS model.\vskip8pt

The VSL framework, therefore, offers a significant shift in perspective:\emph{
rather than supporting a $\Lambda$CDM universe undergoing late-time
acceleration, the Hubble diagram of SNeIa should be reinterpreted
as evidence for a declining speed of light in an expanding Einstein--de
Sitter universe.}

\section{\label{sec:BDRS}Rethinking $\,$Blanchard--Douspis--Rowan-Robinson--Sarkar's
2003 CMB analysis and $\normalsize{\textbf{{\,H$_0\approx\ $46\,}}}$}

Let us now turn our discussion to a remarkable proposal advanced by
Blanchard, Douspis, Rowan-Robinson, and Sarkar (BDRS) in 2003 and
its relation to our SNeIa analysis.\vskip4pt

It is well established that the $\Lambda$CDM model, \emph{augmented}
by the primordial fluctuation spectrum (presumably arising from inflation)
in the form $P(k)=A\,k^{n}$, successfully accounts for the observed
anisotropies in the cosmic microwave background (CMB). This model
predicts a dark energy density of $\Omega_{\Lambda}\approx0.7$, a
Hubble constant of $H_{0}\approx67$, and a spectral index $n\approx0.96$
\citep{CMB-Planck1,CMB-Planck2}. Yet, in \citep{BDRS-2003} BDRS
reanalyzed the CMB, available at the time from the Wilkinson Microwave
Anisotropy Probe (WMAP), in a new perspective. These authors deliberately
relied on the EdS model, which corresponds to a flat $\Lambda$CDM
model with $\Omega_{\Lambda}=0$. Rather than invoking the $\Lambda$
component, they adopted a slightly modified form for the primordial
fluctuation spectrum. They reasoned that, since the spectral index
$n$ is scale-dependent for any polynomial potential of the inflaton
and is constant only for an exponential potential, it is reasonable
to consider a double-power form for the spectrum of primordial fluctuations
\begin{equation}
P(k)=\begin{cases}
A_{1}\,k^{n_{1}} & k\leqslant k_{*}\\
A_{2}\,k^{n_{2}} & k\geqslant k_{*}
\end{cases}
\end{equation}
with a continuity condition ($A_{1}\,k_{*}^{n_{1}}=A_{2}\,k_{*}^{n_{2}}$)
across the breakpoint $k_{*}$.\vskip4pt

Using this new function, BDRS produced an excellent fit to the CMB
power spectrum, resulting in the following parameters: $H_{0}=46$
km/s/Mpc, $\omega_{\text{baryon}}:=\Omega_{\text{baryon}}(H_{0}/100)^{2}=0.019$,
$\tau=0.16$ (the optical depth to last scattering), $k_{*}=0.0096$
Mpc$^{-1}$, $n_{1}=1.015$, and $n_{2}=0.806$. The most remarkable
outcome of the BDRS work is the ``low'' value of $H_{0}=46$, representing
a 34\% reduction from the accepted value of $H_{0}\sim70$. A detailed
follow-up study by Hunt and Sarkar \citep{Hunt-2007}, based on a
supergravity-induced multiple inflation scenario, yielded a comparable
value of $H_{0}\approx44$. Notably, around the same time, Shanks
argued that a value of $H_{0}\lesssim50$ might permit a simpler inflationary
model with $\Omega_{\text{baryon}}=1$, i.e. without invoking dark
energy or cold dark matter \citep{Shanks-2004}.\vskip4pt

The success achieved by BDRS in reproducing the CMB power spectrum
can be interpreted as indicating a degeneracy in the parameter space
$\{\Omega_{\Lambda},\,H_{0}\}$. Specifically, the BDRS pair $\{\Omega_{\Lambda}=0,\,H_{0}=46\}$
is `nearly degenerate' with the canonical pair $\{\Omega_{\Lambda}\approx0.7,\,H_{0}\approx70\}$,
insofar as the CMB data is concerned. Importantly, BDRS's modest modification
in the primordial fluctuation spectrum can make $\Omega_{\Lambda}$
redundant. In other words, \emph{the $\Lambda$ component is vulnerable
to other exogenous underlying assumptions that supplement the $\Lambda$CDM
model}.\vskip4pt

Notably, strong degeneracies in the parameter space related to the
CMB have been reported recently. In \citep{Perivolaropoulos-2020}
Alestas et al found that the best-fit value of $H_{0}$ obtained from
the CMB power spectrum is \emph{degenerate} with a constant equation
of state (EoS) parameter $w$; the relationship is approximately linear,
given by $H_{0}+30.93\,w-36.47=0$ (with $H_{0}$ in km/s/Mpc). Although
this finding is not directly related to the BDRS work, the $H_{0}$-vs-$w$
degeneracy reinforces the general conclusion regarding the sensitivity
of $H_{0}$ to \emph{other} exogenous underlying assumptions that
supplement the $\Lambda$CDM model---in the case of Alestas et al,
the EoS parameter $w$.\vskip4pt

While a drastically low value of $H_{0}\approx46$ at first seems
to be `a steep price to pay', we have demonstrated in the preceding
sections that this new value is fully compatible with the $H_{0}=47.2$
obtained from the Hubble diagram of SNeIa data when analyzed within
the context of VSL cosmology. Consequently, \emph{the $\Lambda$ component
becomes redundant not only for the CMB but also for SNeIa.} \vskip8pt

The alignment of our findings with those of BDRS is especially remarkable
for several reasons:\pagebreak
\begin{itemize}
\item The Hubble diagram of SNeIa and the CMB power spectrum are two ``orthogonal''
datasets. SNeIa data relates to observations along the time direction,
while the CMB captures a two-dimensional snapshot across space at
the recombination event. Furthermore, they correspond to two separate
epochs---one representing late time (SNeIa) and the other representing
early time (CMB)--- each characterized by distinct relevant physics.
\item There is no a priori reason to expect the double-power primordial
fluctuation spectrum used in the BDRS work to result in a reduction
in $H_{0}$ rather than an enhancement. Moreover, there is no inherent
indication of the 34\% change in $H_{0}$. The strength of our VSL
analysis of SNeIa is in its capability to explain both the direction
and magnitude of the change in $H_{0}$ through the $3/2$--factor
in the modified Hubble law; see Section \ref{subsec:H0-reduction}.
\item Our VSL framework is inspired from theoretical consideration of scale-invariant
actions (see Section \ref{sec:Generating-VSL} herein and Ref. \citep{VSL2024-dilaton})
and does not rely on prior knowledge of BDRS's analysis. It was not
deliberately designed to address BDRS's surprise finding of $H_{0}\approx46$.
In this regard, our findings should be viewed as a \emph{retrodiction}
of BDRS's results, supporting $H_{0}\sim46\text{--}47$ and bypassing
the need for the $\Lambda$ component.
\end{itemize}
Together with our SNeIa analysis, the work of BDRS eliminates the
need for the $\Lambda$ component regarding the two `orthogonal' datasets---the
CMB and SNeIa. Future applications of our VSL framework to gravitational
lensing, Baryonic Acoustic Oscillations (BAO), and other areas are
worthwhile.
\begin{figure}[!t]
\begin{centering}
\hskip-6pt\includegraphics[scale=0.6]{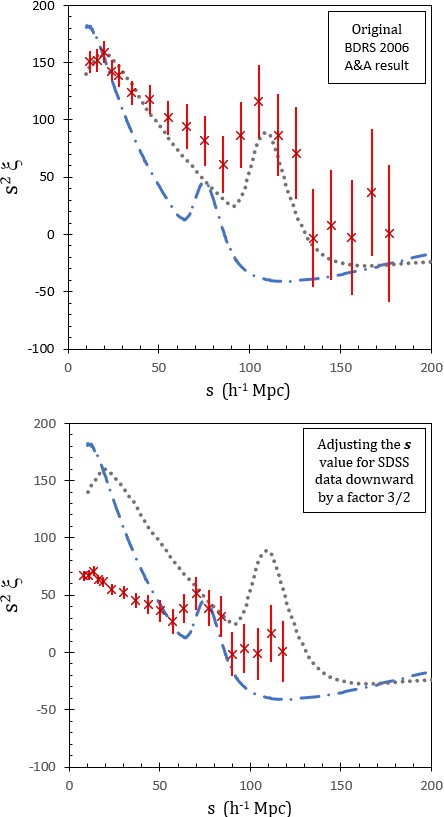}
\par\end{centering}
\caption{The correlation function in observed (redshift) space, as reproduced
from BDRS's 2006 SDSS study \citep{BDRS-2006}. Upper panel: BDRS's
original result. Lower panel: the SDSS data are corrected by reducing
$s$ and $s^{2}\,\xi$ by factors of $3/2$ and $(3/2)^{2}$, respectively.}

\noindent \label{fig:BDRS-2006}
\end{figure}

\subsection*{What caused BDRS to abandon their $H_{0}\approx46$ finding?}

In 2006 BDRS revisited their 2003 A\&A proposal by applying it to
the Sloan Digital Sky Survey (SDSS) of luminous red galaxies (LRG)
which became available in \citep{Eisenstein-2006}. In their 2006
follow-up work \citep{BDRS-2006}, BDRS claimed that the ``low'' value
of $H_{0}\approx46$ was unable to produce an acceptable fit to the
two-point correlation function of LRG in observed (redshift) space.
The upper panel of Fig.$\ $\ref{fig:BDRS-2006} reproduces their
finding, showing that the SDSS data (and their error bars in red segments)
largely aligns with the $\Lambda$CDM model (dotted line), while the
BDRS model (dashed-dotted line) is significantly off. \emph{This discrepancy
eventually forced BDRS to abandon their 2003 proposal in its entirety}
(although Hunt and Sarkar continued with their follow-up study shortly
thereafter \citep{Hunt-2007}).\vskip4pt

However, we believe that BDRS's 2006 SDSS analysis contained an oversight,
in light of our VSL cosmology. The standard Lema\^itre redshift formula
and the conventional Hubble law are not applicable in the presence
of varying speed of light. As discussed in Sections \ref{subsec:Modified-Lemaitre}
and \ref{subsec:Modified-Hubble}, these expressions are modified
by a factor of $3/2$ due to varying $c$. Therefore, the SDSS would
need a reevaluation to incorporate this VSL-induced adjustment. Here,
we tentatively make a rudimentary fix: $\,$we correct the comoving
distance $s$ (measured in multiples of $h$, defined as $H_{0}/(100$
km/s/Mpc)), downward by a factor of $3/2$.\linebreak In the lower
panel of Fig. \ref{fig:BDRS-2006}, we adjust the SDSS data (and their
error bars) by reducing $s$ by a factor of $3/2$ and $s^{2}\,\xi$
by a factor of $(3/2)^{2}$. Upon these adjustments, the peak (at
$s\simeq75$) and trough (at $s\simeq60$) of the SDSS become aligned
with those of the BDRS's model (dashed-dotted line), thereby lessening
the discrepancy issue that led to BDRS's abandonment of their original
proposal.\vskip4pt

We conclude that it was \emph{premature} for BDRS to abandon their
2003 CMB study and the finding of $H_{0,}\approx46$. Rather, their
proposal should be revived and applied to the upgraded Planck dataset
for the CMB \citep{CMB-Planck1}. We should also note that since the
CMB data is a two-dimensional snapshot of the sky at the time at recombination,
VSL is not expected to be a dominant player in the CMB. Nevertheless,
potential impacts of VSL on the CMB are an interesting avenue for
future research.

\section{\label{sec:Resolve-age-prob}Resolving the age problem}

From the definition of the Hubble constant, $H(t):=\frac{1}{a}\frac{da}{dt}$,
and the evolution, $a\propto t^{2/3}$, the age of an EdS universe
is related to the current-time $H_{0}$ value by
\begin{equation}
t_{0}^{\text{EdS}}=\frac{2}{3\,H_{0}}\label{eq:age-formula}
\end{equation}
A value of $H_{0}\sim70$, would result in an age of 9.3 billion years
which would be too short to accommodate the existence of the oldest
stars---a paradox commonly referred to as the age problem.\vskip4pt

Standard cosmology resolves the age problem by invoking the $\Lambda$
component which induces an acceleration phase following a deceleration
phase. The spatially flat $\Lambda$CDM model is known to give the
age formula in an analytical form (with $\Omega_{M}+\Omega_{\Lambda}=1$
and $\Omega_{\Lambda}>0$) \citep{Rubakov-book}\vspace{-.1cm}
\begin{equation}
t_{0}^{\Lambda\text{CDM}}=\frac{2}{3\sqrt{\Omega_{\Lambda}}H_{0}}\,\text{arcsinh}\sqrt{\frac{\Omega_{\Lambda}}{\Omega_{M}}}\,.\label{eq:age-LCDM}
\end{equation}
which restores Eq. \eqref{eq:age-formula} when $\Omega_{M}\rightarrow1$
and $\Omega_{\Lambda}\rightarrow0$. For positive $\Omega_{\Lambda}$,
the age exceeds $2/(3H_{0})$. With $H_{0}=70.2$, $\Omega_{M}=0.285$,
$\Omega_{\Lambda}=0.715$, it yields an age of $13.6$ billion years,
an accepted figure in standard cosmology.\vskip4pt

However, our VSL cosmology naturally overcomes the age problem without
invoking the $\Lambda$ component. The reason is that $H_{0}$ is
reduced by a factor of $3/2$, as detailed in Section \ref{subsec:H0-reduction}.
The \emph{reduced} value $H_{0}=47.22\pm0.4$ (95\% CL) promptly yields
$t_{0}=13.82\pm0.11$ billion years (95\% CL), consistent with the
accepted age value, thereby successfully resolving the age problem.

\section{\label{sec:Resolving-H0-tension}Toward a new resolution of the $\normalsize{\textbf{\,H$_0$\,}}$
tension}

Galaxies are gravitationally bound structures, stabilized by gravitational
attraction and rotational motion of matter within them. However, they
are embedded in a cosmic background that is not static, but rather
expanding over time. As such, stable galaxies in principle may adjust
to the growth in the scale of the ``ambient'' intergalactic space
surrounding them; viz., their local scales may increase in response
to the cosmic expansion. This growth in the local scales of galaxies---if
it exists---would be of \emph{astronomical} nature. To investigate
this phenomenon, one could explore the evolution of a spinning disc-shaped
distribution of matter (serving as a simplified model for a galaxy)
on an expanding cosmic background within the scale-invariant theory
mentioned in Section \ref{sec:Generating-VSL}, although such an exploration
lies beyond the scope of this paper. We should note that recent observational
studies have reported evidence of galaxies experiencing growth in
size \citep{Ormerod-2024,Buitrago-2024}.\vskip4pt

For our purposes, in Section \ref{sec:Refractive}, we have modeled
the local scale $\hat{a}$ of an individual galaxy as a function of
its redshift $z$, supplemented with a negligible idiosyncratic component
that randomly varies from one galaxy to another. The function $F(z)$,
defined in Eq. \eqref{eq:def-F} as the ratio of the local scale of
galaxies at redshift $z$ to the local scale of the Milky Way, captures
the evolution of the local scale over cosmic time. In Section \ref{subsec:Enabling-F-bin},
$F(z)$ was empirically extracted from the Pantheon Catalog, with
Fig. \ref{fig:F-vs-z-and-a} displaying $F(z)$ as a function of the
redshift and of cosmic scale factor, respectively. In accordance with
our expectation, the local scale $\hat{a}$ of galaxies gradually
increases in response to the growth of the global cosmic factor $a$
over cosmic time.

\subsection*{A `running' $\small\text{H$_0$}(z)$ }

\begin{figure}[!t]
\begin{centering}
\hskip-6pt\includegraphics[scale=0.8]{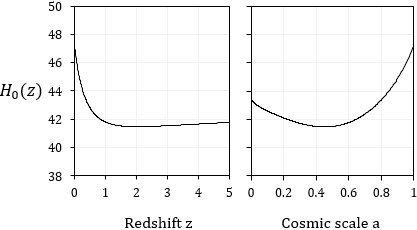}
\par\end{centering}
\caption{The variation of $H_{0}(z)$ as functions of redshift (left panel)
and of cosmic scale factor (right panel).}

\label{fig:H0-vs-z-and-a}
\end{figure}

The function $F(z)$ can be absorbed into an ``effective'' Hubble
constant $H_{0}(z)$ which depends on redshift $z$.\linebreak Specifically,
Formula \eqref{eq:modified-dL-vs-z} can be rewritten as
\begin{equation}
\frac{\hat{d}_{L}}{\hat{c}_{ob}}=\frac{1+z}{\frac{3}{2}H_{0}(z)}\,\ln(1+z)\label{eq:dL-vs-z-sim}
\end{equation}
where the newly introduced function $H_{0}(z)$ is given by
\begin{equation}
H_{0}(z)=H_{0}\,F(z)\,\biggl(1-\frac{3}{2}\frac{\ln F(z)}{\ln(1+z)}\biggr)\,.\label{eq:H0-vs-z}
\end{equation}
Formulae \eqref{eq:dL-vs-z-sim} and \eqref{eq:H0-vs-z} thus allow
for a \emph{current-time} $H_{0}(z)$ `running' as a function of the
redshift of the data that are used to estimate it. With the function
$F(z)$ parametrized in Eq. \eqref{eq:func-form-of-F} with $b_{1}=b_{2}=2$
and $F_{\infty}=0.931$ as produced in Section \ref{subsec:Enabling-F-analytic},
$H_{0}(z)$ can be computed using Eq. \eqref{eq:H0-vs-z}, as displayed
in Fig. \ref{fig:H0-vs-z-and-a}. At first, $H_{0}(z)$ decreases
from $47.2$ (at $z=0$) to $41.5$ (at $z\simeq2$), experiencing
a 12\% reduction. For $z\gtrsim2$, $H_{0}(z)$ slowly rerises. At
$z\rightarrow\infty$, with $F_{\infty}=0.931$ and $H_{0}(z=0)=47.22$,
per Eq. \eqref{eq:H0-vs-z}, $H_{0}(z)$ asymptotically approaches
$F_{\infty}H_{0}=43.96$, representing an 7\% reduction from $H_{0}(z=0)$.\vskip8pt

Two immediate remarks can be made:
\begin{enumerate}
\item Interestingly, the overall 7\% reduction in the $H_{0}$ estimate
at highest-$z$ SNeIa data is of the comparable magnitude with the
discrepancy in $H_{0}$ reported in standard cosmology, which observes
a decreases of $H_{0}$ from 73 (using SNeIa) to 67 (using the Planck
CMB), an 8\% reduction.
\item Remarkably, the asymptotic value $H_{0}(z\rightarrow\infty)=43.96$
that we just derived agrees surprisingly well with the $H_{0}\approx44$
value obtained by Hunt and Sarkar in their follow-up study of the
CMB \citep{Hunt-2007}.
\end{enumerate}
The `running' phenomenon of $H_{0}(z)$ arises because astronomical
objects---either the CMB or SNeIa---are subject to their local scale
which gradually grows during the cosmic expansion.

\noindent 
\vspace{-.5cm}

\subsection*{Hints at an astronomical origin of the $\small\text{H$_0$}$ tension}

We have, therefore, linked the `running' current-time $H_{0}(z)$
with the function $F(z)$. Since $F(z)$ captures the evolution of
galaxies' local scales in response to the growth of the global scale
of intergalactic space, the `running' $H_{0}(z)$ is thus of \emph{astronomical}
origin. The empirical evaluation for $F(z)$ from the Pantheon Catalog,
as detailed in Sections \ref{subsec:Enabling-F-bin} and \ref{subsec:Enabling-F-analytic},
demonstrates that the local scale gradually increases with cosmic
time, indicating that galaxies \emph{cannot} fully resist cosmic expansion.
As mentioned earlier, understanding the growth in $F(z)$ would require
an in-depth examination of a spinning disc-shaped distribution of
matter in an expanding cosmic background within a scale-invariant
theory of gravity and matter, a task that is left for future investigation.
\vskip8pt

We should also note similar works along this line of `running' $H_{0}(z)$
\citep{Krishnan-2020,Krishnan-2021,Dainotti-2021}. For example, in
\citep{Dainotti-2021}, Dainotti et al considered an extension of
the flat $w_{0}w_{a}CDM$. They proposed the following luminosity
distance--redshift formula
\begin{align}
\frac{d_{L}}{c} & =(1+z)\times\nonumber \\
 & \int_{0}^{z}\frac{dz'}{H_{0}(z')\sqrt{\Omega_{0M}(1+z')^{3}+\Omega_{0\Lambda}\,e^{3\int_{0}^{z'}du\frac{1+w(u)}{1+u}}}}
\end{align}
with the parametrization $H_{0}(z)=\tilde{H}_{0}\,(1+z)^{-\alpha}$
and an evolutionary equation of state for the $\Lambda$ component
$w(z)=w_{0}+w_{a}\,z/(1+z)$. In this formula, $H_{0}(z)$ can be
interpreted as a `running' current-time Hubble value, which depends
on the redshift of the data used to estimate it. These authors are
able to bring the value of $H_{0}$ at $z=$1,100 within $1\sigma$
of the Planck measurements, hence effectively removing the $H_{0}$
tension. However, unlike our approach, where the function $F(z)$
has a well-defined\linebreak \emph{astrophysical} interpretation,
the use of $H_{0}(z)$ and $w(z)$ in Ref. \citep{Dainotti-2021}
represents ad hoc parametrizations, with their underlying nature remaining
unknown. Additionally, the $H_{0}(z)$ in Ref. \citep{Dainotti-2021}
is likely of a cosmic origin, whereas the equation of state $w(z)$
of the $\Lambda$ component is of a field theoretical origin.\vskip8pt

In closing of this section, our study offers a potential resolution
to the $H_{0}$ tension. Furthermore, it suggests that this tension
has an astronomical origin, arising from the growth in the local scale
of gravitationally-bound galaxies over cosmic time.

\section{\label{sec:Summary}Discussions and summary}

This paper was inspired by three separate lines of development:\vskip8pt

1. In 2003, Blanchard et al (BDRS) proposed a novel CMB analysis that
avoids the $\Lambda$ component \citep{BDRS-2003}. Based solely on
the EdS model (i.e., $\Omega_{\Lambda}=0$) and adopting a double-power
primordial fluctuation spectrum, BDRS achieved an excellent fit to
WMAP's CMB power spectrum. Surprisingly, they obtained a new value
$H_{0}\approx46$, representing a $34\%$ reduction compared to the
accepted value $H_{0}\sim70$ that relies on the flat $\Lambda$CDM
model with $\Omega_{\Lambda}\approx0.7$.\vskip4pt

As independently reported more recently in \citep{Perivolaropoulos-2020},
there exists a strong degeneracy inherent in the parameter space concerning
the CMB data. Drawn from this observation, we can interpret BDRS's
findings as indicating that \emph{within the flat $\Lambda$CDM model,
the parameter pairs $\{\Omega_{\Lambda}=0,\,H_{0}\approx46\}$ and
$\{\Omega_{\Lambda}\approx0.7,\,H_{0}\approx70\}$ are `nearly degerenate'
insofar as the CMB power spectrum is concerned.} With a modest modification
to the primordial fluctuation spectrum, the BDRS parameter pair \emph{$\{\Omega_{\Lambda}=0,\,H_{0}\approx46\}$}
becomes advantageous over the $\Lambda$CDM pair \emph{$\{\Omega_{\Lambda}\approx0.7,\,H_{0}\approx70\}$}.
While the cost of this modification is not prohibitive, as BDRS provided
justifications in support of a double-power primordial fluctuation
spectrum, the benefit is profound in that the DE hypothesis is rendered
unnecessary.\vskip4pt

This perspective raises an intriguing possibility that the parameter
pairs $\{\Omega_{\Lambda}=0,\,H_{0}\approx46\}$ and $\{\Omega_{\Lambda}\approx0.7,\,H_{0}\approx70\}$
may also be `nearly degenerate' insofar as \emph{the Hubble diagram
of SNeIa }is concerned. To materialize this possibility, one must
first seek an \emph{alternative} approach to late-time acceleration
that does not invoke DE. We should emphasize that such an alternative---if
it exists---must not only eliminate the role of $\Omega_{\Lambda}$
but also reduce the $H_{0}$ value from $\sim70$ to $\sim46$. This
presents a stringent requirement to be met.\vskip8pt

2. A recent theoretical approach developed by the present author \citep{VSL2024-dilaton}
induces a variation in the speed of light $c$ (and a variation in
the quantum of action $\hbar$) from a dynamical dilaton $\chi$.
The derivation applies to a class of scale-invariant actions that
allow matter to couple with the dilaton. The dynamics of $c$ (and
$\hbar$) stems parsimoniously from the dilaton, rather than as serving
as an auxiliary addition to the action. The dependencies are determined
to be $c\propto\chi^{1/2}$ and $\hbar\propto\chi^{-1/2}$. It was
also found that the timescale $\tau$ and lengthscale $l$ of a given
physical process are related in the an anisotropic fashion, $\tau\propto l^{\,3/2}$.
See Section \ref{sec:Generating-VSL}.\vskip8pt

3. Existing efforts in the literature to apply variable speed of light
(VSL) theories to the Hubble diagram of SNeIa have been impeded by
a detrimental oversight. All available VSL analyses to date have relied
on the standard Lema\^itre redshift relation, $1+z=a^{-1}$, leading
to a flawed consensus that VSL plays no role in late-time acceleration.
This error stems from the assumption that $c$ is solely a function
of cosmic time $t$, overlooking the possibility that $c$ can vary
across the boundaries of galaxies, where a gravitationally-bound galactic
region merges with the expanding intergalactic space surrounding it.\vskip8pt

In this paper, we build upon the VSL theory referenced in Point \#2,
correct the error mentioned in Point \#3, and reanalyze the Pantheon
Catalog. The effects of VSL modify the Lema\^iitre formula to $1+z=a^{-3/2}$,
with the $3/2$--exponent arising from the anisotropic time scaling
referenced earlier, $\tau\propto l^{\,3/2}$. Intuitively, this factor
3/2 influences the evaluation of $H_{0}$, resulting in a reduction
from the canonical value of $H_{0}\sim70$ by a factor of $3/2$ to
$H_{0}=47.2$. The new value is compatible with BDRS's findings for
the CMB mentioned in Point \#1.\vskip8pt

\textbf{Our derivation and analysis:} The logical steps of our work
are as follows.\vskip8pt

\paragraph{\emph{(i) }Modifying the FLRW metric.}

The universe is modeled as an EdS spacetime supporting a varying $c$
as (see Eq. \eqref{eq:modified-RW})
\[
\begin{cases}
\ ds^{2} & =\ c^{2}(a)\,dt^{2}-a^{2}(t)\,\left[dr^{2}+r^{2}d\Omega^{2}\right]\\
\ c(a) & =\ c_{0}\left(\frac{a}{a_{0}}\right)^{-1/2}
\end{cases}
\]
with the expansion obeying the growth law $a(t)=a_{0}\bigl(t/t_{0}\bigr)^{2/3}$,
see Eq. \eqref{eq:a-vs-t}. Justifications for this model are provided
in Section \ref{subsec:Modified-RW}.\vskip8pt

\paragraph{\emph{(ii) }Modifying the Lema\^itre redshift formula.}

Across the boundaries of galaxies, $c$ also varies, leading to a
refraction on the lightwaves. Due to this effect, we find that the
classic Lema\^itre redshift formula $1+z=a^{-1}$ is inapplicable
for the VSL cosmology, and is replaced by the \emph{modified} Lema\^itre
redshift formula (see Eq. \eqref{eq:modified-Lemaitre})
\[
1+z=a^{-3/2}\,F^{3/2}(z)
\]
with a new exponent of $3/2$ and $F(z)$ measuring the relative change
in the local scale of galaxies. See Sections \ref{subsec:Frequency-shift},
\ref{sec:Refractive} and \ref{subsec:Modified-Lemaitre}.\vskip8pt

\paragraph{\emph{(iii) }Modifying the Hubble law.}

The $3/2$-exponent in the \emph{modified} Lema\^itre redshift formula
above leads to the \emph{modified} Hubble law (see Eq. \eqref{eq:modified-hubble-law})
\[
z=\frac{3}{2}H_{0}\,\frac{d}{c_{0}}
\]
This new Hubble law differs from the classic Hubble law by a multiplicative
factor of $3/2$, resulting in a reduction in the $H_{0}$ estimate
by a factor of $3/2$. See Section \ref{subsec:Modified-Hubble}.\vskip8pt

\paragraph{\emph{(iv) }Modifying the luminosity distance-vs-\emph{z} formula:}

This formula is the centerpiece of our study (see Eq. \eqref{eq:modified-dL-vs-z})
\[
\frac{\hat{d}_{L}}{\hat{c}_{ob}}=\frac{1+z}{H_{0}F(z)}\,\ln\frac{(1+z)^{2/3}}{F(z)}
\]
See Sections \ref{subsec:Modified-d-z} and \ref{subsec:Modified-dL-z}
for derivation.\vskip8pt

\paragraph{\emph{(v) }A re-analysis of the Pantheon data based on VSL:}

In Section \ref{sec:Reanalysis-Pantheon}, we apply the Formulae above
to the Combined Pantheon Sample of SNeIa. We produce an excellent
fit without invoking the $\Lambda$ component; the fit is as robust
as that obtained from the $\Lambda CDM$ model. The optimal values
for the parameters are:
\begin{itemize}
\item The Hubble constant $H_{0}=47.22\pm0.4$ (95\% CL). This value of
consistent with the $3/2$ reduction referenced in Point (iii) above.
\item The local scale of galaxies decreases with respect to redshift as
$F(z)=1-(1-F_{\infty}).\left(1-(1+z)^{-2}\right)^{2}$, with $F_{\infty}=0.931\pm0.11$
(95\% CL).
\end{itemize}
\vskip8pt

Our modified Lema\^itre redshift formula, Eq. \eqref{eq:modified-Lemaitre},
can also effectively viewed as a form of ``redshift remapping'', a
technique advocated in Refs. \citep{Wojtak-2016,Wojtak-2017,Bassett-2013}.
Interestingly, our value of $H_{0}=47.2\pm0.4$ aligns with the result
$H_{0}=48\pm2$ reported in \citep{Wojtak-2017}.\vskip12pt

\textbf{Implications:} Four important findings emerge from our analysis.
\vskip12pt

\textbf{I)}\textbf{\emph{ Declining speed of light as an alternative
interpretation of the Hubble diagram of SNeIa}}. At high redshift,
the luminosity distance in an EdS universe behaves as $d_{L}\propto z$,
whereas in our VSL cosmology, it behaves as $d_{L}\propto z\,\ln z$.
Due to VSL, high-redshift SNeIa thus benefit from the additional $\ln z$
term, making them appear dimmer than predicted by the EdS model. Another
intuitive way to understand this behavior is to note that since $c\propto a^{-1/2}$,
light traveled faster in the past than in later epochs. As a result,
lightwaves from distant SNeIa require less time to traverse the earlier
sections of their trajectories, hence experiencing less cosmic expansion
(and redshift) than the EdS model predicts. Hence, the high-$z$ section
of the Hubble diagram of SNeIa can be explained---qualitatively and
quantitatively---\emph{by a declining speed of light rather than
a recent cosmic acceleration}. A detailed exposition is given in Section
\ref{sec:VSL-as-alternative}.\textcolor{red}{\vskip8pt}

\textbf{II) }\textbf{\emph{Reviving BDRS's work on the CMB, avoiding
dark energy.}} Despite the very different natures of the data involved,
our VSL-based analysis of SNeIa and BDRS's work on the CMB fully agree
on two aspects: (i) the universe obeys the EdS model (i.e. $\Omega_{\Lambda}=0$),
and (ii) $H_{0}$ is reduced to $46\text{--}47$. The BDRS parameter
pair $\{\Omega_{\Lambda}=0,\,H_{0}\approx46\}$ is advantageous over
the $\Lambda$CDM pair $\{\Omega_{\Lambda}\approx0.7,\,H_{0}\approx70\}$
regarding both the CMB \emph{and} SNeIa, which are `orthogonal' datasets.
Detailed discussions are presented in Section \ref{sec:BDRS}. Together,
our current work and BDRS's 2003 analysis challenge the existence
of dark energy---one of the foundational assumptions of the cosmological
concordance model.\textcolor{red}{\vskip8pt}

\textbf{III)}\textbf{\emph{ Resolving the age problem.}} The age of
an EdS universe is given by: $t_{0}=2/(3H_{0})$. Using the \emph{reduced}
value of $H_{0}=47.22\pm0.4$ km/s/Mpc, one obtains $t_{0}=13.82\pm0.11$
billion years. The age problem is thus resolved through the reduction
in $H_{0}$, without requiring a recent acceleration phase induced
by dark energy. See Section \ref{sec:Resolve-age-prob}.\vskip8pt

\textbf{IV)}\textbf{\emph{ Addressing the $H_{0}$ tension.}} Utilizing
the function $F(z)$, we recast the current-time Hubble constant as
a function $H_{0}(z)$ of redshift. Between $z=0$ and $z\rightarrow\infty$,
the `running' $H_{0}(z)$ exhibits a 7\% decrease, a reduction in
similar magnitude with the ongoing $H_{0}$ tension between the CMB
and SNeIa. See Section \ref{sec:Resolving-H0-tension}.

\subsection*{On the cosmological time dilation extracted from the Dark Energy
Survey (DES)}

A recent paper \citep{DES-Collaboration-2024} using DES supernova
light curves showed no deviation from the relation $\Delta t_{obs}=\Delta t_{em}(1+z)$.
However, this finding does not contradict our modified Lemaitre redshift
formula, $1+z=a^{-3/2}F(z)$. This is because the result in Ref. \citep{DES-Collaboration-2024}
only verifies that the speed of light \emph{inside} the galaxies hosting
the supernovae and that \emph{inside} the Milky Way are approximately
the same. Galaxies are gravitationally bound and thus not subject
to cosmic expansion. Reference \citep{DES-Collaboration-2024} does
not deal with the speed of light in the intergalactic space, the expansion
of which causes c to decline over time. We clarified this distinction
in Section \ref{sec:Refractive}.

\section{Conclusion}

The nearly identical agreement of the CMB and SNeIa regarding the
reduced value of $H_{0}\sim46\text{--}47$ is highly encouraging.
This alignment points toward a consistent cosmological framework based
on the Einstein--de Sitter model with a variable speed of light,
thus eliminating the need for dark energy and dissolving its fine-tuning
and coincidence problems.\vskip8pt

Importantly, we have built a case for an alternative perspective:
\emph{$\,$rather than supporting a $\Lambda$CDM universe undergoing
late-time acceleration, the Hubble diagram of SNeIa can be reinterpreted
as evidence for a declining speed of light in an expanding Einstein--de
Sitter universe}.\vskip8pt

Finally, we note that the observational bounds established in the
literature in support of a constant speed of light have predominantly
relied on standard cosmology \citep{Abdo-2009,Agrawal-2021,Cai-2016,Cao-2017,Colaco-2022,Qi-2014,Ravanpak-2017,Salzano-2016a,Salzano-2016b,Santos-2024,Uzan-2003,Uzan-2011,Wang-2019,Zhang-2014,Cao-2018,Zou-2018,Mendonca-2021}.
However, our new Lema\^itre redshift formula represents a critical
departure from this conventional framework. Therefore, the consensus
regarding the absence of variation in $c$ in observational cosmology
must be reconsidered in light of our findings, prompting a comprehensive
reanalysis of these constraints.
\begin{acknowledgments}
I thank Clifford Burgess, Tiberiu Harko, Robert Mann, Anne-Christine
Davis, Eoin \'O Colg\'ain, Demosthenes Kazanas, Leandros Perivolaropoulos,
John Moffat, Alberto Salvio, Ilya Shapiro, Gregory Volovik and Bruce
Bassett for their constructive feedback.
\end{acknowledgments}

\noindent \begin{center}
\vskip10pt---------------$\infty$---------------
\par\end{center}

\appendix

\section{\label{sec:Refraction-effect}$\ $Refraction effect}

\begin{figure}[!t]
\begin{centering}
\includegraphics[scale=0.7]{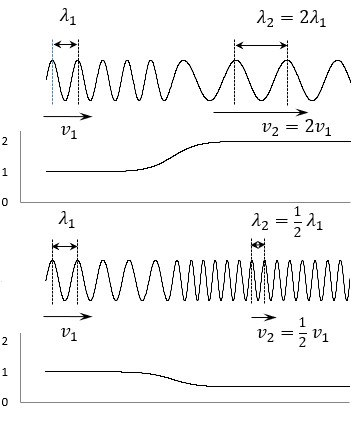}
\par\end{centering}
\caption{Change in wavelength as a wavetrain travels in a medium with varying
speed of wave. Upper panel: wavelength doubles as its speed doubles.
Lower panel: wavelength halves as its speed halves. In either case,
wavelength and speed are proportional: $\lambda_{2}/v_{2}=\lambda_{1}/v_{1}$.}

\label{fig:wavetrains}
\end{figure}
Let us start with a well-understood phenomenon: the behavior of a
wavetrain in a medium with varying speed of wave. It is well established
that the wavelength of the wavetrain at a given location is proportional
to the speed of wave at that location:

\noindent 
\begin{equation}
\lambda\propto v\label{eq:lambda-v-speed-1}
\end{equation}
Figure \ref{fig:wavetrains} illustrates the change in wavelength
as a wave travels at varying speed. In the upper panel, as the speed
increases, the front end of the wavecrest will rush forward leaving
its back end behind thus stretching out the wavecrest. In the lower
panel, the reverse situation occurs: as the speed decreases, the front
end of the wavecrest will slow down while its back end continues its
course thus compressing the wavecrest. In either situation, the wavelength
and the speed of wave are directly proportional:
\begin{equation}
\frac{\lambda_{2}}{\lambda_{1}}=\frac{v_{2}}{v_{1}}\label{eq:lambda-v-speed-2}
\end{equation}
Note that the details of how the variation of $v$ does \emph{not}
participate in formula above.

\section{\label{sec:Equivalent-derivation}$\ $An equivalent derivation of
the modified Lema\^itre redshift formula}

We produce an alternative route by way of frequency transformation
to modifying Lema\^itre's redshift formula \eqref{eq:modified-Lemaitre}.
We have derived in Eq. \eqref{eq:freq_shift} that
\begin{equation}
\frac{\nu_{ob}}{\nu_{em}}=\frac{a_{em}^{3/2}}{a_{ob}^{3/2}}\label{eq:freq_shift-1}
\end{equation}

\noindent For transits between local regions to global regions (i.e.,
Transit \#1 and Transit \#3 in Fig. \ref{fig:wavetrain-full} in Page
\pageref{fig:wavetrain-full}), since $\lambda\propto c$, the frequency
is:
\begin{align}
\nu & =\frac{c}{\lambda}=\text{const }
\end{align}

\noindent This means that the frequency of the lightwave does not
change during Transit \#1 and Transit \#3, viz.
\begin{align}
\hat{\nu}_{em} & =\nu_{em};\ \ \ \hat{\nu}_{ob}=\nu_{ob}
\end{align}
Given that
\begin{align}
\hat{\lambda}_{ob} & =\frac{\hat{c}_{ob}}{\hat{\nu}_{ob}};\ \ \ \hat{\lambda}_{em}=\frac{\hat{c}_{em}}{\hat{\nu}_{em}};\ \ \ \frac{\lambda^{*}}{\hat{\lambda}_{em}}=\frac{\hat{a}_{ob}}{\hat{a}_{em}}
\end{align}
and
\begin{equation}
\frac{\hat{c}_{ob}}{\hat{c}_{em}}=\frac{\hat{a}_{ob}^{-1/2}}{\hat{a}_{em}^{-1/2}}
\end{equation}
it is straightforward to verify that
\begin{align}
1+z & =\frac{\hat{\lambda}_{ob}}{\lambda^{*}}=\frac{a_{ob}^{3/2}}{a_{em}^{3/2}}\,\frac{\hat{a}_{em}^{3/2}}{\hat{a}_{ob}^{3/2}}
\end{align}
a relation that is in perfect agreement with Eq. \eqref{eq:modified-Lemaitre}.


\begin{thebibliography}{100}
\bibitem{Perivolaropoulos-2022}L. Perivolaropoulos and F. Skara,
\emph{Challenges for \textgreek{L}CDM: An update}, New Astron. Rev.
\textbf{95}, 101659 (2022), \textcolor{purple}{\href{https://arxiv.org/abs/2105.05208}{arXiv:2105.05208 [astro-ph.CO]}}

\bibitem{diValentino-2021}E. Di Valentino, O. Mena, S. Pan, L. Visinelli,
W. Yang, A. Melchiorri, D. F. Mota, A. G. Riess and J. Silk,\emph{
In the realm of the Hubble tension---a review of solutions}, Class.
Quant. Grav. 38 (2021) no.15, 153001, \textcolor{purple}{\href{https://arxiv.org/abs/2103.01183}{arXiv:2103.01183 [astro-ph.CO]}}

\bibitem{Bull}P. Bull, Y. Arkrami, et al,\emph{ Beyond \textgreek{L}CDM:
Problems, solutions, and the road ahead}, Physics of the Dark Universe
12 (2016) 56, \textcolor{purple}{\href{https://arxiv.org/abs/1512.05356}{arXiv:1512.05356}}

\bibitem{BDRS-2003}A. Blanchard, M. Douspis, M. Rowan-Robinson and
S. Sarkar, \emph{An alternative to the cosmological \textquotedblleft concordance
model\textquotedblright }, Astron. Astrophys. \textbf{412}, 35-44
(2003), \textcolor{purple}{\href{https://arxiv.org/abs/astro-ph/0304237}{arXiv:astro-ph/0304237}}

\bibitem{Hunt-2007}P. Hunt and S. Sarkar, \emph{Multiple inflation
and the WMAP \textquoteleft \textquoteleft glitches\textquoteright \textquoteright .
II. Data analysis and cosmological parameter extraction}, Phys. Rev.
D\textbf{ 76}, 123504 (2007), \textcolor{purple}{\href{https://arxiv.org/abs/0706.2443}{arXiv:0706.2443 [astro-ph]}}

\bibitem{VSL2024-dilaton}H. K. Nguyen, \emph{Dilaton-induced variations
in Planck constant and speed of light: An alternative to Dark Energy},
Phys. Lett. B \textbf{862} (2025) 139357, \textcolor{purple}{\href{https://arxiv.org/abs/2412.04257}{arXiv:2412.04257 [gr-qc]}}

\bibitem{VSL2024-1}H. K. Nguyen, \emph{A mechanism to generate varying
speed of light via Higgs-dilaton coupling: Theory and cosmological
applications}, Eur. Phys. J. C \textbf{85}, 393 (2025), \textcolor{purple}{\href{https://arxiv.org/abs/2408.02583}{arXiv:2408.02583 [gr-qc]}}

\bibitem{Einstein-1911}A. Einstein, \emph{On the influence of gravitation
on the propagation of light,} Annalen der Physik \textbf{35}, 898-908
(1911), \textcolor{purple}{\href{https://einsteinpapers.press.princeton.edu/vol3-trans/393}{Link to English translation}}

\bibitem{Einstein-1912-a}A. Einstein, \emph{The speed of light and
the statics of the gravitational field,} Annalen der Physik \textbf{38},
355-369 (1912), \textcolor{purple}{\href{https://einsteinpapers.press.princeton.edu/vol4-trans/107}{Link to English translation}}

\bibitem{Einstein-1912-b}A. Einstein, \emph{Relativity and gravitation:
Reply to a comment by M. Abraham,} Annalen der Physik \textbf{38},
1059-1064 (1912), \textcolor{purple}{\href{https://einsteinpapers.press.princeton.edu/vol4-trans/142}{Link to English translation}}

\bibitem{Dicke-1957}R. H. Dicke, \emph{Gravitation without a Principle
of Equivalence}, Rev. Mod. Phys. \textbf{29}, 363 (1957) 

\bibitem{BransDicke-1961}C. H. Brans and R. Dicke, \emph{Mach's Principle
and a relativistic theory of gravitation}, Phys. Rev. \textbf{124},
925 (1961)

\bibitem{Moffat-1993}J. W. Moffat, \emph{Superluminary universe:
A possible solution to the initial value problem in cosmology}, Int.
J. Mod. Phys. D \textbf{2}, 351 (1993), \textcolor{purple}{\href{https://arxiv.org/abs/gr-qc/9211020}{arXiv:gr-qc/9211020}}

\bibitem{Albrecht-1998}A. Albrecht and J. Magueijo, \emph{Time varying
speed of light as a solution to cosmological puzzles,} Phys. Rev.
D \textbf{59}, 043516 (1999), \textcolor{purple}{\href{https://arxiv.org/abs/astro-ph/9811018}{arXiv:astro-ph/9811018}}

\bibitem{Barrow-1998a}J. D. Barrow, \emph{Cosmologies with varying
light speed}, Phys. Rev. D \textbf{59}, 043515 (1999), \textcolor{purple}{\href{https://arxiv.org/abs/astro-ph/9811022}{arXiv:astro-ph/9811022}}

\bibitem{Barrow-1998b}J. D. Barrow and J. Magueijo, \emph{Varying-$\alpha$
theories and solutions to the cosmological problems}, Phys. Lett.
B \textbf{443}, 104 (1998), \textcolor{purple}{\href{https://arxiv.org/abs/astro-ph/9811072}{arXiv:astro-ph/9811072}}

\bibitem{Barrow-2000}J. D. Barrow and J. Magueijo, \emph{Can a changing
\textgreek{a} explain the Supernovae results?}, Astrophys. J. \textbf{532},
L87--L90 (2000), \textcolor{purple}{\href{https://arxiv.org/abs/astro-ph/9907354}{arXiv:astro-ph/9907354}}

\bibitem{Magueijo-2002}J. Magueijo and L. Smolin, \emph{Lorentz invariance
with an invariant energy scale}, Phys. Rev. Lett. \textbf{88}, 190403
(2002), \textcolor{purple}{\href{https://arxiv.org/abs/hep-th/0112090}{arXiv:hep-th/0112090}}

\bibitem{Magueijo-2003}J. Magueijo, \emph{New varying speed of light
theories}, Rept. Prog. Phys. \textbf{66}, 2025 (2003), \textcolor{purple}{\href{https://arxiv.org/abs/astro-ph/0305457}{arxiv:astro-ph/0305457}}

\bibitem{Barrow-1999a}J. D. Barrow and J. Magueijo, \emph{Solutions
to the Quasi-flatness and Quasi-lambda Problems}, Phys. Lett. B \textbf{447},
246 (1999), \textcolor{purple}{\href{https://arxiv.org/abs/astro-ph/9811073}{arXiv:astro-ph/9811073}}

\bibitem{Barrow-1999b}J. D. Barrow and J. Magueijo, \emph{Solving
the Flatness and Quasi-flatness Problems in Brans-Dicke Cosmologies
with a Varying Light Speed}, Class. Quant. Grav.\textbf{ 16}, 1435
(1999), \textcolor{purple}{\href{https://arxiv.org/abs/astro-ph/9901049}{arXiv:astro-ph/9901049}}

\bibitem{Clayton-1999}M. A. Clayton and J. W. Moffat, \emph{Dynamical
mechanism for varying light velocity as a solution to cosmological
problems}, Phys. Lett. B 460, 263 (1999), \textcolor{purple}{\href{https://arxiv.org/abs/astro-ph/9812481}{arXiv:astro-ph/9812481 [astro-ph]}}

\bibitem{Avelino-1999}P. P. Avelino and C. J. A. P. Martins, \emph{Does
a varying speed of light solve the cosmological problems?}, Phys.
Lett. B \textbf{459} (1999), 468-472, \textcolor{purple}{\href{https://arxiv.org/abs/astro-ph/9906117}{arXiv:astro-ph/9906117 [astro-ph]}}

\bibitem{Clayton-2000}M. A. Clayton and J. W. Moffat, \emph{Scalar
tensor gravity theory for dynamical light velocity}, Phys. Lett. B
\textbf{477} (2000), 269-275, \textcolor{purple}{\href{https://arxiv.org/abs/gr-qc/9910112}{arXiv:gr-qc/9910112 [gr-qc]}}

\bibitem{Magueijo-2000}J. Magueijo, \emph{Covariant and locally Lorentz
invariant varying speed of light theories}, Phys. Rev. D \textbf{62}
(2000), 103521, \textcolor{purple}{\href{https://arxiv.org/abs/gr-qc/0007036}{arXiv:gr-qc/0007036}}

\bibitem{Clayton-2002}M. A. Clayton and J. W. Moffat,\emph{ Vector
field mediated models of dynamical light velocity}, Int. J. Mod. Phys.
D \textbf{11} (2002), 187-206, \textcolor{purple}{\href{https://arxiv.org/abs/gr-qc/0003070}{arXiv:gr-qc/0003070 [gr-qc]}}

\bibitem{Bassett-2000}B. A. Bassett, S. Liberati, C. Molina-Paris
and M. Visser, \emph{Geometrodynamics of Variable-Speed-of-Light Cosmologies},
Phys. Rev. D\textbf{ 62}, 103518 (2000), \textcolor{purple}{\href{https://arxiv.org/abs/astro-ph/0001441}{arXiv:astro-ph/0001441}}

\bibitem{Liberati-2000}S. Liberati, B. A. Bassett, C. Molina-Paris
and M. Visser, \emph{Chi-Variable-Speed-of-Light Cosmologies}, Nucl.
Phys. Proc. Suppl. \textbf{88}, 259 (2000), \textcolor{purple}{\href{https://arxiv.org/abs/astro-ph/0001481}{arXiv:astro-ph/0001481}}

\bibitem{Drummond-1999}I. T. Drummond, \emph{Variable Light-Cone
Theory of Gravity}, \textcolor{purple}{\href{https://arxiv.org/abs/gr-qc/9908058}{arXiv:gr-qc/9908058}}

\bibitem{Drummond-1980}I. T. Drummond and S. J. Hathrell, \emph{QED
vacuum polarization in a background gravitational field and its effect
on the velocity of photons}, Phys. Rev. D \textbf{22}, 343 (1980)

\bibitem{Novello-1989}M. Novello and S. D. Jorda, \emph{Does there
exist a cosmological horizon problem?}, Mod. Phys. Lett. A \textbf{4},
1809 (1989)

\bibitem{Volovik-2023}G. E. Volovik, \emph{Planck constants in the
symmetry breaking quantum gravity}, Symmetry \textbf{15}, 991 (2023),
\textcolor{purple}{\href{https://arxiv.org/abs/2304.04235}{arXiv:2304.04235 [cond-mat.other]}}

\bibitem{Zhang-2014}P. Zhang and X. Meng, \emph{SNe data analysis
in variable speed of light cosmologies without cosmological constant},
Mod. Phys. Lett. A Vol. \textbf{29}, No. 24, 1450103 (2014), \textcolor{purple}{\href{https://arxiv.org/abs/1404.7693}{arXiv:1404.7693 [astro-ph.CO]}}

\bibitem{Qi-2014}J-Z. Qi, M-J. Zhang and W-B. Liu, \emph{Observational
constraint on the varying speed of light theory}, Phys. Rev. D \textbf{90},
063526 (2014), \textcolor{purple}{\href{https://arxiv.org/abs/1407.1265}{arXiv:1407.1265 [gr-qc]}}

\bibitem{Ravanpak-2017}A. Ravanpak, H. Farajollahi and G.F. Fadakar,
\emph{Normal DGP in varying speed of light cosmology}, Res. Astron.
Astrophys. \textbf{17}, 26 (2017), \textcolor{purple}{\href{https://arxiv.org/abs/1703.09811}{arXiv:1703.09811}}

\bibitem{Salzano-2016a}V. Salzano and M. P. Dabrowski, \emph{Statistical
hierarchy of varying speed of light cosmologies}, Astrophys. J. \textbf{851},
97 (2017), \textcolor{purple}{\href{https://arxiv.org/abs/1612.06367}{arXiv:1612.06367}}

\bibitem{Salzano-2016b}A. Balcerzak, M. P. Dabrowski and V. Salzano,
\emph{Modelling spatial variations of the speed of light}, Annalen
der Physik \textbf{29}, 1600409 (2017), \textcolor{purple}{\href{https://arxiv.org/abs/1604.07655}{arXiv:1604.07655 [astro-ph.CO]}}

\bibitem{Gupta-2020}R. P. Gupta, \emph{Cosmology with relativistically
varying physical constants}, Mon. Not. Roy. Astron. Soc. \textbf{498}
(2020) 3, 4481-4491, \textcolor{purple}{\href{https://arxiv.org/abs/2009.08878}{arXiv:2009.08878 [astro-ph.CO]}}

\bibitem{Gupta-2021}R. P. Gupta, \emph{Varying physical constants
and the lithium problem}, Astroparticle Physics \textbf{129}, 102578
(2021), \textcolor{purple}{\href{https://arxiv.org/abs/2010.13628}{arXiv:2010.13628 [gr-qc]}}

\bibitem{Cuzinatto-2022}R. R. Cuzinatto, R. P. Gupta, R. F. L. Holanda,
J. F. Jesus and S. H. Pereira, \emph{Testing a varying-\textgreek{L}
model for dark energy within Co-varying Physical Couplings framework},
Mon. Not. Roy. Astron. Soc. \textbf{515}, 5981-5992 (2022), \textcolor{purple}{\href{https://arxiv.org/abs/2204.10764}{arXiv:2204.10764 [gr-qc]}}

\bibitem{Abdo-2009}A. A. Abdo et al, \emph{A limit on the variation
of the speed of light arising from quantum gravity effects}, Nature
\textbf{462}, 331-334 (2009)

\bibitem{Agrawal-2021}R. Agrawal, H. Singirikonda and S. Desai, \emph{Search
for Lorentz Invariance Violation from stacked Gamma-Ray Burst spectral
lag data}, JCAP\textbf{ 05} (2021) 029, \textcolor{purple}{\href{https://arxiv.org/abs/2102.11248}{arXiv:2102.11248 [astro-ph.HE]}}

\bibitem{Santos-2024}J. Santos, C. Bengaly, B. J. Morais and R. S.
Goncalves, \emph{Measuring the speed of light with cosmological observations:
current constraints and forecasts}, JCAP \textbf{11} (2024) 062, \textcolor{purple}{\href{https://arxiv.org/abs/2409.05838}{arXiv:2409.05838 [astro-ph.CO]}}

\bibitem{Uzan-2003}J. P. Uzan, \emph{The Fundamental Constants and
Their Variation: Observational Status and Theoretical Motivations},
Rev. Mod. Phys. \textbf{75} (2003), 403, \textcolor{purple}{\href{https://arxiv.org/abs/hep-ph/0205340}{arXiv:hep-ph/0205340 [hep-ph]}}

\bibitem{Uzan-2011}J. P. Uzan, \emph{Varying Constants, Gravitation
and Cosmology}, Living Rev. Rel. \textbf{14} (2011), 2, \textcolor{purple}{\href{https://arxiv.org/abs/1009.5514}{arXiv:1009.5514 [astro-ph.CO]}}

\bibitem{Martins-2017}C. J. A. P. Martins, \emph{The status of varying
constants: a review of the physics, searches and implications}, Rep.
Prog. Phys. \textbf{80}, 126902 (2017), \textcolor{purple}{\href{https://arxiv.org/abs/1709.02923}{arXiv:1709.02923 [astro-ph.CO]}}

\bibitem{Ellis-2005}G. F. R. Ellis and J. P. Uzan, \emph{\textquoteleft c\textquoteright{}
is the speed of light, isn\textquoteright t it?}, Am. J. Phys. 73
(2005), 240-247, \textcolor{purple}{\href{https://arxiv.org/abs/gr-qc/0305099}{arXiv:gr-qc/0305099 [gr-qc]}}

\bibitem{Ellis-2007}G. F. R. Ellis, \emph{Note on Varying Speed of
Light Cosmologies}, Gen. Rel. Grav. \textbf{39} (2007), 511-520, \href{https://arxiv.org/abs/astro-ph/0703751}{arXiv:astro-ph/0703751 [astro-ph]}
\textcolor{purple}{\href{https://arxiv.org/abs/astro-ph/0703751}{https://arxiv.org/abs/astro-ph/0703751}}

\bibitem{Magueijo-2008}J. Magueijo and J. W. Moffat, \emph{Comments
on \textquotedbl Note on varying speed of light theories\textquotedbl},
Gen. Rel. Grav. \textbf{40}, 1797-1806 (2008), \textcolor{purple}{\href{https://arxiv.org/abs/0705.4507}{arXiv:0705.4507}}

\bibitem{Cruz-2012}C. N. Cruz and A. C. A. de Faria Jr., \emph{Variation
of the speed of light with temperature of the expanding universe},
Phys. Rev. D \textbf{86} (2012), 027703, \textcolor{purple}{\href{https://arxiv.org/abs/1205.2298}{arXiv:1205.2298 [gr-qc]}}

\bibitem{Moffat-2016}J. W. Moffat, \emph{Variable Speed of Light
Cosmology, Primordial Fluctuations and Gravitational Waves}, Eur.
Phys. J. C \textbf{76}, 130 (2016), \textcolor{purple}{\href{https://arxiv.org/abs/1404.5567}{arXiv:1404.5567 [astro-ph.CO]}}

\bibitem{Franzmann-2017}G. Franzmann, \emph{Varying fundamental constants:
a full covariant approach and cosmological applications}, \textcolor{purple}{\href{https://arxiv.org/abs/1704.07368}{arXiv:1704.07368 [gr-qc]}}

\bibitem{Cruz-2018}C. N. Cruz and F. A. da Silva, \emph{Variation
of the speed of light and a minimum speed in the scenario of an inflationary
universe with accelerated expansion}, Phys. Dark Univ. \textbf{22}
(2018), 127-136, \textcolor{purple}{\href{https://arxiv.org/abs/2009.05397}{arXiv:2009.05397 [physics.gen-ph]}}

\bibitem{Costa-2019}R. Costa, R. R. Cuzinatto, E. M. G. Ferreira
and G. Franzmann, \emph{Covariant c-flation: a variational approach},
Int. J. Mod. Phys. D \textbf{28}, 1950119 (2019), \textcolor{purple}{\href{https://arxiv.org/abs/1705.03461}{arXiv:1705.03461 [gr-qc]}}

\bibitem{Lee-2021a}S. Lee, \emph{The minimally extended Varying Speed
of Light (meVSL)}, JCAP \textbf{08} (2021), 054, \textcolor{purple}{\href{https://arxiv.org/abs/2011.09274}{arXiv:2011.09274 [astro-ph.CO]}}

\bibitem{Balcerzak-2014a}A. Balcerzak and M. P. Dabrowski, \emph{Redshift
drift in varying speed of light cosmology}, Phys. Lett. B \textbf{728}
(2014), 15, \textcolor{purple}{\href{https://arxiv.org/abs/1310.7231}{arXiv:1310.7231 [astro-ph.CO]}}

\bibitem{Balcerzak-2014b}A. Balcerzak and M. P. Dabrowski, \emph{A
statefinder luminosity distance formula in varying speed of light
cosmology}, JCAP \textbf{06} (2014), 035, \textcolor{purple}{\href{https://arxiv.org/abs/1406.0150}{arXiv:1406.0150 [astro-ph.CO]}}

\bibitem{Salzano-2015}V. Salzano, M. P. Dabrowski and R. Lazkoz,
\emph{Measuring the speed of light with Baryon Acoustic Oscillations,}
Phys. Rev. Lett. \textbf{114}, 101304 (2015), \textcolor{purple}{\href{https://arxiv.org/abs/1412.5653}{arXiv:1412.5653 [astro-ph.CO]}}

\bibitem{Salzano-2016c}V. Salzano, M. P. Dabrowski and R. Lazkoz,
\emph{Probing the constancy of the speed of light with future galaxy
survey: The case of SKA and Euclid}, Phys. Rev. D \textbf{93}, 063521
(2016), \textcolor{purple}{\href{https://arxiv.org/abs/1511.04732}{arXiv:1511.04732 [astro-ph.CO]}}

\bibitem{Cai-2016}R. G. Cai, Z. K. Guo and T. Yang, \emph{Dodging
the cosmic curvature to probe the constancy of the speed of light},
JCAP \textbf{08} (2016), 016, \textcolor{purple}{\href{https://arxiv.org/abs/1601.05497}{arXiv:1601.05497 [astro-ph.CO]}}

\bibitem{Cao-2017}S. Cao, M. Biesiada, J. Jackson, X. Zheng, Y. Zhao
and Z. H. Zhu, \emph{Measuring the speed of light with ultra-compact
radio quasars}, JCAP \textbf{02} (2017), 012, \textcolor{purple}{\href{https://arxiv.org/abs/1609.08748}{arXiv:1609.08748 [astro-ph.CO]}}

\bibitem{Salzano-2017a}V. Salzano, \emph{How to Reconstruct a Varying
Speed of Light Signal from Baryon Acoustic Oscillations Surveys},
Universe 3 (2017) no.2, 35

\bibitem{Salzano-2017b}V. Salzano, \emph{Recovering a redshift-extended
VSL signal from galaxy surveys}, Phys. Rev. D \textbf{95}, 084035
(2017), \textcolor{purple}{\href{https://arxiv.org/abs/1604.03398}{arXiv:1604.03398 [astro-ph.CO]}}

\bibitem{Lang-2018}R. G. Lang, H. Martinez-Huerta and V. de Souza,
\emph{Limits on the Lorentz Invariance Violation from UHECR astrophysics},
Astrophys. J. \textbf{853} (2018) no.1, 23, \textcolor{purple}{\href{https://arxiv.org/abs/1701.04865}{arXiv:1701.04865 [astro-ph.HE]}}

\bibitem{Zou-2018}X. B. Zou, H. K. Deng, Z. Y. Yin and H. Wei, \emph{Model-Independent
Constraints on Lorentz Invariance Violation via the Cosmographic Approach},
Phys. Lett. B \textbf{776} (2018), 284-294, \textcolor{purple}{\href{https://arxiv.org/abs/1707.06367}{arXiv:1707.06367 [gr-qc]}}

\bibitem{HWAC-2018}H. Martinez-Huerta {[}HAWC{]}, \emph{Potential
constrains on Lorentz invariance violation from the HAWC TeV gamma-rays},
PoS ICRC2017 (2018), 868, \textcolor{purple}{\href{https://arxiv.org/abs/1708.03384}{arXiv:1708.03384 [astro-ph.HE]}}

\bibitem{Cao-2018}S. Cao, J. Qi, M. Biesiada, X. Zheng, T. Xu and
Z. H. Zhu, \emph{Testing the Speed of Light over Cosmological Distances:
The Combination of Strongly Lensed and Unlensed Type Ia Supernovae},
Astrophys. J. \textbf{867} (2018) no.1, 50, \textcolor{purple}{\href{https://arxiv.org/abs/1810.01287}{arXiv:1810.01287 [astro-ph.CO]}}

\bibitem{Wang-2019}D. Wang, H. Zhang, J. Zheng, Y. Wang and G. B.
Zhao, \emph{Reconstructing the temporal evolution of the speed of
light in a flat FRW Universe}, Res. Astron. Astrophys. \textbf{19}
(2019) 10, 152,\textcolor{purple}{{} \href{https://arxiv.org/abs/1904.04041}{arXiv:1904.04041 [astro-ph.CO]}}

\bibitem{HAWC-2020}A. Albert et al. {[}HAWC{]}, \emph{Constraints
on Lorentz Invariance Violation from HAWC Observations of Gamma Rays
above 100 TeV}, Phys. Rev. Lett. \textbf{124} (2020) no.13, 131101,
\textcolor{purple}{\href{https://arxiv.org/abs/1911.08070}{arXiv:1911.08070 [astroph.HE]}}

\bibitem{Pan-2020}Y. Pan, J. Qi, S. Cao, T. Liu, Y. Liu, S. Geng,
Y. Lian and Z. H. Zhu, \emph{Model-independent constraints on Lorentz
invariance violation: implication from updated Gamma-ray burst observations},
Astrophys. J. \textbf{890} (2020), 169, \textcolor{purple}{\href{https://arxiv.org/abs/2001.08451}{arXiv:2001.08451 [astro-ph.CO]}}

\bibitem{Mendonca-2021}I. E. C. R. Mendonca, K. Bora, R. F. L. Holanda,
S. Desai and S. H. Pereira, \emph{A search for the variation of speed
of light using galaxy cluster gas mass fraction measurements}, JCAP
\textbf{11} (2021), 034, \textcolor{purple}{\href{https://arxiv.org/abs/2109.14512}{arXiv:2109.14512 [astro-ph.CO]}}

\bibitem{Rodrigues-2022}G. Rodrigues and C. Bengaly, \emph{A model-independent
test of speed of light variability with cosmological observations},
JCAP \textbf{07} (2022) no.07, 029, \textcolor{purple}{\href{https://arxiv.org/abs/2112.01963}{arXiv:2112.01963 [astro-ph.CO]}}

\bibitem{Lee-2023a}S. Lee, \emph{Constraining minimally extended
varying speed of light by cosmological chronometers}, Mon. Not. Roy.
Astron. Soc. \textbf{522} (2023) no.3, 3248-3255, \textcolor{purple}{\href{https://arxiv.org/abs/2301.06947}{arXiv:2301.06947 [astro-ph.CO]}}

\bibitem{Eaves-2022}R. E. Eaves, \emph{Redshift in varying speed
of light cosmology}, Mon. Not. Roy. Astron. Soc. \textbf{516}, 4136--4145
(2022)

\bibitem{Mukherjee-2024}P. Mukherjee, G. Rodrigues and C. Bengaly,
\emph{Examining the validity of the minimal varying speed of light
model through cosmological observations: Relaxing the null curvature
constraint}, Phys. Dark Univ. \textbf{43} (2024), 101380, \textcolor{purple}{\href{https://arxiv.org/abs/2302.00867}{arXiv:2302.00867 [astro-ph.CO]}}

\bibitem{Liu-2023}Y. Liu, S. Cao, M. Biesiada, Y. Lian, X. Liu and
Y. Zhang, \emph{Measuring the Speed of Light with Updated Hubble Diagram
of High-redshift Standard Candles}, Astrophys. J. \textbf{949} (2023)
no.2, 57, \textcolor{purple}{\href{https://arxiv.org/abs/2303.14674}{arXiv:2303.14674 [astro-ph.CO]}}

\bibitem{Lee-2023b}S. Lee, \emph{Constraint on the minimally extended
varying speed of light using time dilations in Type Ia supernovae},
Mon. Not. Roy. Astron. Soc. \textbf{524} (2023) no.3, 4019-4023, \textcolor{purple}{\href{https://arxiv.org/abs/2302.09735}{arXiv:2302.09735 [astro-ph.CO]}}

\bibitem{Lee-2021b}S. Lee, \emph{Constraints on the time variation
of the speed of light using Strong lensing}, \textcolor{purple}{\href{https://arxiv.org/abs/2104.09690}{arXiv:2104.09690 [astro-ph.CO]}}

\bibitem{Cuzinatto-2023}R. R. Cuzinatto, C. A. M. de Melo and J.
C. S. Neves, \emph{Shadows of black holes at cosmological distances
in the co-varying physical couplings framework}, Mon. Not. Roy. Astron.
Soc. \textbf{526} (2023) no.3, 3987-3993, \textcolor{purple}{\href{https://arxiv.org/abs/2305.11118}{arXiv:2305.11118 [gr-qc]}}

\bibitem{Zhang-2024}C. Y. Zhang, W. Hong, Y. C. Wang and T. J. Zhang,
\emph{A Stochastic Approach to Reconstructing the Speed of Light in
Cosmology}, Mon. Not. Roy. Astron. Soc. \textbf{534} (2024), 56-69,
\textcolor{purple}{\href{https://arxiv.org/abs/2409.03248}{arXiv:2409.03248 [astro-ph.CO]}}

\bibitem{Liu-2021}T. Liu, S. Cao, M. Biesiada, Y. Liu, Y. Lian and
Y. Zhang, \emph{Consistency testing for invariance of the speed of
light at different redshifts: the newest results from strong lensing
and Type Ia supernovae observations}, Mon. Not. Roy. Astron. Soc.
\textbf{506} (2021) 2, 2181-2188, \textcolor{purple}{\href{https://arxiv.org/abs/2106.15145}{arXiv:2106.15145 [astro-ph.CO]}}

\bibitem{Colaco-2022}L. R. Colaco, S. J. Landau, J. E. Gonzalez,
J. Spinelly and G. L. F. Santos, \emph{Constraining a possible time-variation
of the speed of light along with the fine-structure constant using
strong gravitational lensing and Type Ia supernovae observations},
JCAP\textbf{ 08} (2022) 062, \textcolor{purple}{\href{https://arxiv.org/abs/2204.06459}{arXiv:2204.06459 [astro-ph.CO]}}

\bibitem{Liu-2018}Y. Liu and B-Q. Ma, \emph{Light speed variation
from gamma ray bursts: criteria for low energy photons}, Eur. Phys.
J. C \textbf{78} (2018) 825, \textcolor{purple}{\href{https://arxiv.org/abs/1810.00636}{arXiv:1810.00636 [astro-ph.HE]}}

\bibitem{Xu-2016a}H. Xu and B-Q. Ma, \emph{Light speed variation
from gamma-ray bursts}, Astropart. Phys. \textbf{82}, 72 (2016), \textcolor{purple}{\href{https://arxiv.org/abs/1607.03203}{arXiv:1607.03203 [hep-ph]}}

\bibitem{Xu-2016b}H. Xu and B-Q. Ma, \emph{Light speed variation
from gamma ray burst GRB 160509A}, Phys. Lett. B \textbf{760} (2016)
602-604, \textcolor{purple}{\href{https://arxiv.org/abs/1607.08043}{arXiv:1607.08043 [hep-ph]}}

\bibitem{Zhu-2021}J. Zhu and B-Q. Ma, \emph{Pre-burst events of gamma-ray
bursts with light speed variation}, Phys. Lett. B \textbf{820} (2021)
136518, \textcolor{purple}{\href{https://arxiv.org/abs/2108.05804}{arXiv:2108.05804 [astro-ph.HE]}}

\bibitem{Avelino-2000}P. P. Avelino, C. J. A. P. Martins and G. Rocha,
\emph{VSL theories and the Doppler peak}, Phys. Lett. B \textbf{483}
(2000) 210, \textcolor{purple}{\href{https://arxiv.org/abs/astro-ph/0001292}{arXiv:astro-ph/0001292}}

\bibitem{Buchalter-2004}A. Buchalter, \emph{On the time variation
of c, G, and h and the dynamics of the cosmic expansion}, \textcolor{purple}{\href{https://arxiv.org/abs/astro-ph/0403202}{arXiv:astro-ph/0403202}}

\bibitem{Blas-2011}D. Blas, M. Shaposhnikov and D. Zenhausern, \emph{Scale-invariant
alternatives to general relativity}, Phys. Rev. D\textbf{ 84}, 044001
(2011), \textcolor{purple}{\href{https://arxiv.org/abs/1104.1392}{arXiv:1104.1392 [hep-th]}}

\bibitem{Ferreira-2016}P. G. Ferreira, C. T. Hill and G. G. Ross,
\emph{No fifth force in a scale invariant universe}, Phys. Rev. D
\textbf{95}, 064038 (2017), \textcolor{purple}{\href{https://arxiv.org/abs/1612.03157}{arXiv:1612.03157 [gr-qc]}}

\bibitem{Fujii-1982}Y. Fujii, \emph{Origin of the gravitational constant
and particle masses in scale invariant scalar--tensor theory}, Phys.
Rev. D \textbf{26}, 2580 (1982)

\bibitem{Wetterich-1988a}C. Wetterich, \emph{Cosmologies with variable
Newton's `constant'}, Nucl. Phys. B \textbf{302}, 645 (1988)

\bibitem{Wetterich-1988b}C. Wetterich, \emph{Cosmology and the fate
of dilatation symmetry}, Nucl. Phys. B \textbf{302}, 668 (1988), \textcolor{purple}{\href{https://arxiv.org/abs/1711.03844}{arXiv:1711.03844 [hep-th]}}

\bibitem{Wetterich-2013a}C. Wetterich, \emph{Variable gravity Universe},
Phys. Rev. D \textbf{89}, 024005 (2014), \textcolor{purple}{\href{https://arxiv.org/abs/1308.1019}{arXiv:1308.1019 [astro-ph.CO]}}

\bibitem{Scolnic-2018}D. M. Scolnic et al, \emph{The complete light-curve
sample of spectroscopically confirmed Type Ia supernovae from Pan-STARRS1
and cosmological constraints from the Combined Pantheon Sample}, Astrophys.
J. \textbf{859} (2018) 2, 101, \textcolor{purple}{\href{https://arxiv.org/abs/1710.00845}{arXiv:1710.00845 [astro-ph.CO]}}

\bibitem{Pantheon-data}We used the following data file: archive.stsci.edu/hlsps
/ps1cosmo/scolnic/data\_fitres/hlsp\_ps1cosmo\_panstarrs\_ gpc1\_all\_model\_v1\_ancillary-g10.fitres

\bibitem{Riess-1998}A. Riess et al, \emph{Observational evidence
from supernovae for an accelerating universe and a cosmological constant},
Astron. J. \textbf{116}, 1009 (1998), \textcolor{purple}{\href{https://arxiv.org/abs/astro-ph/9805201}{arXiv:astro-ph/9805201}}

\bibitem{Perlmutter-1999}S. Perlmutter et al, \emph{Measurements
of $\Omega$ and $\Lambda$ from 42 high-redshift supernovae}, Astron.
J. \textbf{517}, 565 (1999), \textcolor{purple}{\href{https://arxiv.org/abs/astro-ph/9812133}{arXiv:astro-ph/9812133}}

\bibitem{CMB-Planck1}P. A. R. Ade et al. {[}Planck{]}, \emph{Planck
2015 results. XIII. Cosmological parameter}s, Astron. Astrophys. \textbf{594}
(2016) A13, \textcolor{purple}{\href{https://arxiv.org/abs/1502.01589}{arXiv:1502.01589 [astro-ph.CO]}}

\bibitem{CMB-Planck2}P. A. R. Ade et al. {[}Planck{]}, \emph{Planck
2015 results. XX. Constraints on inflation}, Astron. Astrophys. \textbf{594}
(2016) A20, \textcolor{purple}{\href{https://arxiv.org/abs/1502.02114}{arXiv:1502.02114 [astro-ph.CO]}}

\bibitem{Shanks-2004}T. Shanks, \emph{Problems with the Current Cosmological
Paradigm}, IAU Symp. \textbf{216}, 398 (2005), \textcolor{purple}{\href{https://arxiv.org/abs/astro-ph/0401409}{arXiv:astro-ph/0401409}}

\bibitem{Perivolaropoulos-2020}G. Alestas, L. Kazantzidis and L.
Perivolaropoulos, \emph{$H_{0}$ Tension, Phantom Dark Energy and
Cosmological Parameter Degeneracies}, Phys. Rev. D \textbf{101}, 123516
(2020), \textcolor{purple}{\href{https://arxiv.org/abs/2004.08363}{arXiv:2004.08363 [astro-ph.CO]}}

\bibitem{Eisenstein-2006}D. J. Eisenstein et al, \emph{Detection
of the Baryon Acoustic Peak in the Large-Scale Correlation Function
of SDSS Luminous Red Galaxies}, Astrophys. J. \textbf{633}, 560 (2005),
\textcolor{purple}{\href{https://arxiv.org/abs/astro-ph/0501171}{arXiv:astro-ph/0501171}}

\bibitem{BDRS-2006}A. Blanchard, M. Douspis, M. Rowan-Robinson and
S. Sarkar, \emph{Large-scale galaxy correlations as a test for dark
energy}, Astron. Astrophys. \textbf{449}, 925 (2006), \textcolor{purple}{\href{https://arxiv.org/abs/astro-ph/0512085}{arXiv:astro-ph/0512085}}

\bibitem{Rubakov-book}E.g., see D. S. Gorbunov and V. A. Rubakov,
\emph{Introduction to the Theory of the Early Universe: Hot big bang
theory}, Word Scientific (2011), page 68

\bibitem{Ormerod-2024}K. Ormerod, C. J. Conselice, N. J. Adams, T.
Harvey, D. Austin, J. Trussler, L. Ferreira, J. Caruana, G. Lucatelli,
Q. Li and W. J. Roper, \emph{EPOCHS VI: the size and shape evolution
of galaxies since $z\sim8$ with JWST Observations}, Mon. Not. Roy.
Astron. Soc. \textbf{527}, 6110--6125 (2024), \textcolor{purple}{\href{https://arxiv.org/abs/2309.04377}{arXiv:2309.04377 [astro-ph.GA]}}

\bibitem{Buitrago-2024}F. Buitrago and I. Trujillo, \emph{Strong
size evolution of disc galaxies since $z=1$: Readdressing galaxy
growth using a physically motivated size indicator}, Astron. Astrophys.
\textbf{682}, A110 (2024), \textcolor{purple}{\href{https://arxiv.org/abs/2311.07656}{arXiv:2311.07656 [astro-ph.GA]}}

\bibitem{Krishnan-2020}C. Krishnan, E. \'O Colg\'ain, Ruchika,
A. A. Sen, M. M. Sheikh-Jabbari and T. Yang, \emph{Is there an early
Universe solution to Hubble tension?}, Phys. Rev. D \textbf{102},
103525 (2020), \textcolor{purple}{\href{https://arxiv.org/abs/2002.06044}{arXiv:2002.06044 [astro-ph.CO]}}

\bibitem{Krishnan-2021}C. Krishnan, E. \'O Colg\'ain, M. M. Sheikh-Jabbari
and T. Yang, \emph{Running Hubble Tension and a $H_{0}$ Diagnostic},
Phys. Rev. D \textbf{103}, 103509 (2021), \textcolor{purple}{\href{https://arxiv.org/abs/2011.02858}{arXiv:2011.02858 [astro-ph.CO]}}

\bibitem{Dainotti-2021}M. G. Dainotti, B. De Simone, T. Schiavone,
G. Montani, E. Rinaldi and G. Lambiase, \emph{On the Hubble constant
tension in the SNe Ia Pantheon sample}, Astrophys. J. \textbf{912}
(2021) 2, 150, \textcolor{purple}{\href{https://arxiv.org/abs/2103.02117}{arXiv:2103.02117 [astro-ph.CO]}}

\bibitem{Bassett-2013}B. A. Bassett, Y. Fantaye, R. Hlo\v{z}ek,
C. Sabiu and M. Smith, \emph{Observational Constraints on Redshift
Remapping}, \textcolor{purple}{\href{https://arxiv.org/abs/1312.2593}{arXiv:1312.2593 [astro-ph.CO]}}

\bibitem{Wojtak-2016}R. Wojtak and F. Prada, \emph{Testing the mapping
between redshift and cosmic scale factor}, Mon. Not. Roy. Astron.
Soc. \textbf{458}, 3331 (2016), \textcolor{purple}{\href{https://arxiv.org/abs/1602.02231}{arXiv:1602.02231 [astro-ph.CO]}}

\bibitem{Wojtak-2017}R. Wojtak and F. Prada, \emph{Redshift remapping
and cosmic acceleration in dark-matter-dominated cosmological models},
Mon. Not. Roy. Astron. Soc. \textbf{470}, 4493 (2017),\textcolor{purple}{{}
\href{https://arxiv.org/abs/1610.03599}{arXiv:1610.03599 [astro-ph.CO]}}

\bibitem{DES-Collaboration-2024}R. M. T. White et al {[}DES Collaboration{]},
\emph{The Dark Energy Survey Supernova Program: slow supernovae show
cosmological time dilation out to }$z\sim1$, Mon. Not. Roy. Astron.
Soc. \textbf{533} (2024) 3, 3365-3378, \textcolor{purple}{\href{https://arxiv.org/abs/2406.05050}{arXiv:2406.05050 [astro-ph.CO]}}
\end{thebibliography}
\end{document}